\documentclass[twocolumn]{aastex701}

\usepackage{wrapfig}
\usepackage{multirow}
\usepackage{amsmath}
\usepackage{tabularx}
\usepackage{seqsplit}

\urldef{\globuslink}\url{https://app.globus.org/file-manager?origin_id=fef57d95-ed3b-430f-a4a3-132a95d83a00}

\begin{document}

\title{Prospects for Near-Far Constraints on Reionization from Simulated LSST Observations of Milky Way Dwarf Galaxies and Stellar Streams}

\correspondingauthor{Aritra Kundu}
\email{aritra@sas.upenn.edu}

\author[0000-0001-8746-4753]{Aritra Kundu}
\affiliation{Department of Physics \& Astronomy, University of Pennsylvania, 209 S 33rd St., Philadelphia, PA 19104, USA}
\email{aritra@sas.upenn.edu}


\author[0000-0003-3939-3297]{Robyn Sanderson}
\affiliation{Department of Physics \& Astronomy, University of Pennsylvania, 209 S 33rd St., Philadelphia, PA 19104, USA}
\email{robynes@sas.upenn.edu}

\author[0000-0003-2497-091X]{Nora Shipp}
\affiliation{Department of Astronomy and DiRAC Institute, University of Washington, 3910 15th Ave NE, Seattle, WA, 98195, USA}
\email{nshipp@uw.edu}

\author[0000-0001-7928-1973]{Adrien Thob}
\affiliation{Department of Physics \& Astronomy, University of Pennsylvania, 209 S 33rd St., Philadelphia, PA 19104, USA}
\email{athob@sas.upenn.edu}

\author[0000-0002-3950-9598]{Adam Lidz}
\affiliation{Department of Physics \& Astronomy, University of Pennsylvania, 209 S 33rd St., Philadelphia, PA 19104, USA}
\email{alidz@sas.upenn.edu}

\author[0000-0001-5214-8822]{Nondh Panithanpaisal}
\affiliation{The Observatories of the Carnegie Institution for Science, 813 Santa Barbara St, Pasadena, CA 91101, USA}
\affiliation{TAPIR, California Institute of Technology, Pasadena, CA 91125, USA}
\email{npanithanpaisal@carnegiescience.edu}

\begin{abstract}

The NSF-DOE Vera C.\ Rubin Observatory's Legacy Survey of Space and Time (LSST) will detect resolved stellar populations of intact dwarf galaxies and disrupted stellar streams in the Milky Way (MW) halo and beyond, opening a new avenue for near-field cosmology, particularly for constraining the faint galaxies during reionization. To interpret these data, we present and publicly release mock LSST catalogs of satellites, stellar streams, phase-mixed galaxies, and field galaxies around 13 MW-mass hosts in the FIRE-2 zoom-in simulations, generated with \textsf{py-ananke}. Using a surface brightness limit of $\mu_{g} \leq 34\,\rm{mag\,arcsec^{-2}}$, we find that $\approx29\%$ of streams with $M_{\star} \gtrsim 5 \times 10^5\,M_{\odot}$ are detectable with tidal tails, while another $\approx38\%$ are detectable only through their bound progenitors and would appear as satellites. We estimate that $\approx96\%$ of satellites and $\approx64\%$ of field galaxies (out to $\sim$3.5\,Mpc) with $M_{\star} \gtrsim 10^{5}\,M_{\odot}$ are detectable, decreasing to $\approx$59\% and $\approx$42\% for $M_{\star} \gtrsim 10^{4.5}\,M_{\odot}$ owing to resolution limits. Including streams detectable as satellites reproduces the radial distribution of observed MW satellites, suggesting that some observed satellites may be bound remnants of disrupted systems. Our mock galaxies match the sizes of observed satellites with $M_{\rm{V}} \lesssim -6$. Star-formation histories computed from only the detectable stars broadly recover the simulated star-formation histories. As an application, we test the near-far technique for reconstructing the proto-MW ultraviolet luminosity function (UVLF) during reionization ($z \gtrsim 6$). Adding the detectable streams improves the recovered faint-end slope at $z = 7$ and reduces the halo-to-halo scatter by $\approx33\%$. 

\end{abstract}



\section{Introduction} \label{sec:intro}

Resolved stellar populations of the Milky Way (MW) and Local Group (LG) provide a unique window into understanding galaxy formation and evolution. Photometric and spectroscopic surveys continue to extend observations of dwarf galaxies to fainter luminosities and enable detailed reconstructions of their star-formation histories (SFHs), establishing them as effective near-field probes of the early Universe \citep{Tolstoy_2009, Boylan-Kolchin_2015, Weisz_2014a, Weisz_2014c, Weisz_2017, Simon_2019}. The ancient stellar populations preserved in these systems carry the fossil record of their high-redshift progenitors at $z \gtrsim 6$, which are thought to have played a key role in reionizing the Universe. The forthcoming Legacy Survey of Space and Time (LSST) at the NSF-DOE Vera C.\ Rubin Observatory \citep{Ivezic_2019} is poised to substantially expand the census of near-field resolved stellar populations by detecting more faint satellites and low-surface-brightness stellar streams in the MW halo, and field galaxies outside its halo, offering an unprecedented opportunity to understand the early Universe through near-field cosmology. In particular, exploiting this window to probe the Epoch of Reionization (EoR) requires understanding both the connection between present-day systems and their high-$z$ progenitors and the limits of the upcoming surveys.

Reionization is a major phase transition in the history of the Universe when the ultraviolet (UV) radiation from the first stars, galaxies, and accreting black holes gradually photoionized the neutral hydrogen in the intergalactic medium (IGM) \citep{Barkana_2001, Furlanetto_2006}. Although the properties of the primary ionizing sources remain uncertain, the most likely scenario is that star-forming galaxies are the main drivers of reionization, with the process completing between $z \sim 5.5$ and $z \sim 6$ \citep{Shapiro_1994, Madau_1999, Faucher-Giguere_2008b, Becker_2013, Becker_2015, Zhu_2021, Bosman_2022, Zhu_2023}. The Hubble Space Telescope (HST) has directly observed  reionization-era galaxies down to $M_{\mathrm{UV}} \sim -13$ in fields behind lensing clusters \citep{Finkelstein_2015, Bouwens_2017, Livermore_2017, Atek_2018, Bouwens_2022a}, and the James Webb Space Telescope (JWST) is extending these measurements to $z \gtrsim 10$ \citep{Finkelstein_2023, Leung_2023, Perez-Gonzalez_2023, Finkelstein_2024, Adams_2024, Harikane_2025} and slightly fainter magnitudes (e.g., $M_{\mathrm{UV}} \sim -12$) \citep{Chemerynska_2025, Atek_2025}.

Characterizing the population of the lowest-mass galaxies at high redshifts, and establishing its connection to the nature and abundance of present-day low-mass galaxies, is central to understanding galaxy formation in the past and present. A significant fraction of the ionizing photon budget at high redshift may come from galaxies beyond the reach of HST or JWST: some reionization models predict that faint yet numerous low-mass galaxies, potentially with $M_{\mathrm{UV}}$ as low as $ \sim -8$, contribute the bulk of the ionizing photons \citep{Kuhlen_2012, Wise_2014, Robertson_2015, Atek_2015, Ma_2018a, Ma_2018b, Finkelstein_2019, Atek_2024, Mascia_2024, Simmonds_2024, Harshan_2024, Wu_2024}. These low-mass systems are not only sensitive probes of the quenching due to reionization \citep{Bullock_2000, Brown_2014, Katz_2020, Wang_2021, McQuinn_2024b, McQuinn_2024c}, but also of the impact of stellar feedback \citep{Wise_2008, Bullock_2017, Kimm_2017, Sales_2022} and the nature of dark matter \citep{Nadler_2021, Sipple_2025}. Furthermore, observations of faint galaxies behind lensing clusters carry significant systematic uncertainties from modeling the cluster mass distributions, intracluster light contamination, and other effects \citep{Atek_2018, Bouwens_2022b}. Even fainter sources are ever more difficult to measure directly, and many of them are expected to remain inaccessible even to JWST \citep{Boylan-Kolchin_2015, Boylan-Kolchin_2016, Weisz_2017}.  Hence, indirect methods are required to constrain the faint end of the high-$z$ UVLF, which is important for quantifying the role of low-luminosity galaxies in reionization, and for understanding galaxy formation and evolution more generally.

The ``near-far'' approach is one such promising avenue that uses the present-day ($z = 0$) MW satellites and LG dwarf galaxies as fossil records of their $z \gtrsim 6$ progenitors \citep{Boylan-Kolchin_2015, Boylan-Kolchin_2016, Weisz_2017, Weisz_2014a}. The color-magnitude diagrams (CMDs) of nearby faint galaxies obtained from resolved HST/JWST photometry allow accurate reconstruction of their SFHs \citep{Weisz_2014a, Weisz_2014b}, from which the UV luminosities of their $z \gtrsim 6$ progenitors can be inferred using stellar population synthesis models \citep{Weisz_2014c, Weisz_2017}. This indirect technique provides access to magnitudes as faint as $M_{\rm UV} \sim -10$, extending below the limits of direct high-$z$ observations.

To date, applications of the near-far technique to probe the faint end of the UVLF at  $z \gtrsim 6$ have been restricted to surviving MW satellites and intact LG dwarf galaxies \citep{Boylan-Kolchin_2016, Weisz_2017, Weisz_2014c, Skillman_2017, Savino_2023, McQuinn_2024a}. However, this ignores the significant population of low-mass galaxies that have been disrupted by galactic tides by the present day  and are now in the form of stellar streams and phase-mixed galaxies \citep{Santistevan_2020, Panithanpaisal_2021, Horta_2024}. In \citet{Kundu_2026}, we used the FIRE-2 simulations to show that these disrupted substructures account for up to $\sim$50\% of the total stellar mass budget of the proto-MW/LG and its environment at $z = 6$--$9$. Their inclusion improves the normalization of the reconstructed high-$z$ UVLF by factors of $\sim$2--3 and reduces the halo-to-halo scatter in the recovered faint-end slope by $\sim$20--40\%. These results indicate that disrupted substructures are crucial for a comprehensive near-far analysis, but the feasibility of incorporating them depends on whether they can be detected in sufficient numbers in current and upcoming surveys, and whether they preserve fossil-record information comparable to that of intact dwarf galaxies.

Disrupted substructures such as stellar streams and phase-mixed 
galaxies are difficult to detect because of their low surface brightness ($\mu \gtrsim 27$ $\rm{mag \,\, arcsec^{-2}}$; \citealt{Johnston_2008}). However, recent wide-area surveys such as the \textit{Gaia} mission \citep{Gaia_2016, Gaia_2018, Gaia_2021}, the Dark Energy Survey (DES) \citep{DES_2015, DES_2018, DES_2021}, the Southern Stellar Stream Spectroscopic Survey \citep[$S^5$;][]{Li_2019}, the Large sky Area Multi-Object fiber Spectroscopic Telescope \citep[LAMOST;][]{Zhao_2012}, and the Apache Point Observatory Galactic Evolution Experiment \citep[APOGEE;][]{Majewski_2017} have discovered a substantial population of streams and amassed a large number of high-quality observations of their stars. The total number of known stellar streams from dwarf galaxies and globular clusters in the MW is now approaching $\sim$100 \citep{Malhan_2022, Mateu_2023}. These detections have advanced our understanding of the MW's assembly history \citep[e.g.,][]{Helmi_1999, Ibata_2001, Newberg_2002, Majewski_2003, Grillmair_2006, Newberg_2009, Balbinot_2016, Shipp_2018, Malhan_2018, Belokurov_2018, Li_2019, Necib_2020, Malhan_2021, Helmi_2018, Haywood_2018, Myeong_2018a, Myeong_2018b, Martin_2022, Johnson_2023}. However, the full accretion history of the MW, encoded in its stellar streams, is expected to extend beyond the sensitivity of these surveys, to surface brightnesses fainter than $30$ $\rm{mag \,\, arcsec^{-2}}$ \citep{Shipp_2023}. With LSST \citep{Ivezic_2019}, we will be able to probe this regime, enabling the detection of substantially fainter substructures within the MW halo and obtaining their resolved CMDs to sufficient depth for reconstruction. Recent predictions suggest that LSST could detect $\sim$89 MW satellite galaxies within its footprint \citep{Tsiane_2025}, up from the $\sim$65 currently known \citep[e.g.,][and references therein]{Pace_2025}, and could significantly increase the number of known stellar streams \citep{Shipp_2023}. Realizing the potential of these new data for the near-far technique requires accurate predictions for what LSST will actually recover. 

In \citet{Kundu_2026}, we presented a preliminary detectability estimate of our simulated intact galaxies and stellar streams, finding that $\sim$50\% of streams passed our simple detectability test, which did not distinguish between streams detectable through their tidal tails and those detectable only through their bound progenitors. In this paper, we extend that work by constructing mock LSST catalogs of both intact satellites and stellar streams from the same simulations, and by estimating their detectability through a more realistic approach. We focus on intact galaxies and stellar streams in this paper and defer the detectability of fully phase-mixed galaxies to future work, since the detection methods differ substantially once a galaxy's stars are fully incorporated into the stellar halo of the host. 

The mock catalogs generated in this work are released publicly alongside this paper, which serves as documentation of the public data. While our analysis is focused on probing the faint-end slope of the UVLF at $z \gtrsim 6$ through the near-far technique, the catalogs have a broad range of applications for studies of hierarchical structure formation, tidal disruption, orbital dynamics, and the nature of dark matter. We describe the FIRE-2 simulations and the classification of present-day simulated galaxies in Section~\ref{sec:simulations}; in Section~\ref{sec:mock_catalogs} we walk through the construction of our mock catalogs. In Section~\ref{sec:detectability} we present the detectability analysis of our mock streams and intact galaxies and highlight their present-day properties in Section~\ref{sec:present-day_props}. We compare the mock and simulated SFHs and analyze them in Section~\ref{sec:SFHs}; in Section~\ref{sec:UVLFs} we present our reconstructed UVLFs and the faint-end slope obtained using the detectable substructures. 

\section{Simulations} \label{sec:simulations}

\subsection{FIRE-2 Simulations} \label{subsec:FIRE-2_sims}

We use cosmological zoom-in baryonic simulations from the Feedback in Realistic Environments (FIRE) project \citep{Wetzel_2023, Wetzel_2025}\footnote{\url{https://fire.northwestern.edu/}} to construct our catalog of present-day intact and disrupted galaxies. The simulations are run with the \textsc{Gizmo} code \citep{Hopkins_2015, Hopkins_2018a}, which uses a meshless finite-mass (MFM) Lagrangian method for accurate hydrodynamics. Baryonic processes such as star formation and stellar feedback are implemented using the FIRE-2 numerical prescriptions \citep{Hopkins_2018a}, with realistic radiative heating and cooling processes in the multiphase interstellar medium (ISM) across temperatures from $10$ $\rm{K}$ to $10^{10}$ $\rm{K}$. Star particles form out of gas parcels satisfying a Jeans instability criterion, and inherit the mass and elemental abundances from their parent gas cells. Each star particle represents a single-age, single-abundance stellar population, whose evolution is based on stellar population synthesis models from \textsc{STARBURST99} \citep{Leitherer_1999} assuming a Kroupa initial mass function (IMF; \citealt{Kroupa_2001}). FIRE-2 models stellar feedback processes including supernovae, radiation pressure, continuous mass loss, photoionization and photoelectric heating in a time-resolved manner. 

The simulations adopt a spatially uniform UV/X-ray ionizing background that evolves with redshift \citep{Faucher-Giguere_2009}. This model has two important deficiencies for our purposes: it neglects the inhomogeneous nature of reionization, which may affect the degree of halo-to-halo scatter \citep{Borrow_2023}, and assumes an earlier reionization redshift ($z \sim 10$) than recent measurements suggest ($z \sim 7$; \citealt{Planck_2020}). However, all simulations used in this work consistently employ this model; hence these assumptions will not introduce any simulation-to-simulation difference in our analysis. Additionally, these simulations include a spurious heating of neutral gas at $\lesssim$1000 K at $z \gtrsim 10$ due to cosmic rays, which suppresses star formation in low-mass halos at high redshifts. After reionization begins ($z \lesssim 9$), however, this has a negligible effect \citep{Gandhi_2024}, so we restrict our analysis to $z \lesssim 9$. For a more detailed description of the physics implemented in FIRE-2 simulations, see \citet{Hopkins_2018a, Hopkins_2018b}. 

The simulations successfully generate isolated MW-like galaxies and MW/M31-like galaxy pairs, reproducing many observed properties of the MW \citep{Ma_2017, Hopkins_2018a, Sanderson_2020, Ansar_2025} and its satellite galaxies \citep{Wetzel_2016, Garrison-Kimmel_2019a, Garrison-Kimmel_2019b, Samuel_2020, Samuel_2021}. 
The central massive MW-like galaxy that forms self-consistently in these simulations efficiently destroys satellite galaxies and low-mass subhalos that orbit within the inner $\sim$30 kpc through gravitational tidal forces \citep{Garrison-Kimmel_2017}, yielding a radial distribution of surviving satellite galaxies consistent with observations \citep{Samuel_2020}. The tidal disruption of well-resolved simulated satellite galaxies also leads to a realistic population of stellar streams in terms of number and mass \citep{Panithanpaisal_2021, Horta_2023, Shipp_2023, Kundu_2026}, although they appear to be distributed at larger radii than the known MW streams \citep{Shipp_2023}. The discrepancy in the radial distribution of \emph{streams} can be partially attributed to resolution effects that can change whether disrupted galaxies in the inner halo are classified as streams or phase-mixed, but not whether they disrupt or not. A higher-resolution halo in FIRE have shown the radial distribution of streams in that halo to be broadly consistent with that of the MW streams \citep{Meziani_2026}. Furthermore, streams with still-bound components are likely to survive longer in higher-resolution simulations than lower-resolution ones \citep{Riley_2025, Meziani_2026}, but they still have tidal tails and are not classified as satellites in the simulations. We will discuss the implications of this finding for our results in Section~\ref{subsec:det_stream_gal_results}. Streams from globular clusters do not naturally form in the FIRE-2 simulations due to limited resolution. 


In \citet{Kundu_2026}, we constructed a catalog of surviving intact galaxies, stellar streams, and phase-mixed galaxies from 13 MW-mass halos in the core FIRE-2 simulations: seven isolated MW-mass halos from the \textit{Latte} suite \citep{Wetzel_2016}, and three LG-like MW/M31 pairs from the \textit{ELVIS on FIRE} suite \citep{Garrison-Kimmel_2019a}, along with their surrounding low-mass galaxies (field dwarfs). All simulations follow a $\Lambda$CDM model based on the cosmological parameters from the \citet{Planck_2020} results. The star/gas particles have an initial mass resolution of $\sim$3500--7100 $M_\odot$ and a spatial resolution of 1--4 pc, while the dark matter (DM) particles have mass resolution of 19000--39000 $M_\odot$ and a spatial resolution of $\sim$40 pc. The main host halos have total masses of $M_{\rm 200m}$ $\approx$1--2$\times$ 10$^{12}$ $M_\odot$\footnote{`200m' indicates a measurement relative to 200 times the mean matter density of the Universe.} and stellar masses of $M_\star \approx 10^{10\text{--}11}$ $M_\odot$, spanning the measured values for the MW \citep{Sanderson_2020}. For each simulation, cosmological zoom-in initial conditions are generated at $z$ = 99 from the \textsc{MUSIC} code \citep{Hahn_2011}, with the zoomed regions selected within periodic cosmological boxes of sizes 70--172 Mpc. In each simulation, 600 snapshots are saved from $z = 99$ to $z = 0$, with a typical spacing of $\sim$ 25 Myr after $z\sim 2$, enabling accurate tracking of disrupted galaxies.

\subsection{Classification of Present-day Galaxies in the Simulations} \label{subsec:classify_galaxies}

Using the halo catalogs based on the \textsc{Rockstar} 6D phase-space halo finder \citep{Behroozi_2013a} and the merger trees constructed using \textsc{consistent-trees} \citep{Behroozi_2013b}, we track each subhalo (and its star particles) that falls within the virial radius of the main host galaxy, starting from a high redshift ($z_{{\rm MR}_{3:1}}$) to the present day. This is the redshift at which the stellar-mass ratio of the most massive halo to the second-most massive halo is 3:1, which is when the proto-host galaxy emerges \citep{Santistevan_2020, Horta_2024}. We refer the interested reader to \citet{Necib_2019, Samuel_2020, Gandhi_2024, Kundu_2026} for discussions about the halo finding criteria and star particle assignments at both low- and high-$z$. For more details about the algorithm used to track the galaxies, we refer to \citet{Panithanpaisal_2021, Kundu_2026, Mandacaru-Guerra_2026}. We classify the present-day galaxies into four categories:



\begin{enumerate}
    \item \textbf{Surviving satellite galaxies:} Intact low-mass galaxies within the virial radius (radii)\footnote{In this work, we consider the virial radius to be $R_{\rm 200m}$, which is the spherical radius within which the matter density is 200 times the mean matter density of the Universe.}  of the isolated (paired) MW-like (MW/M31-like) host(s) at $z = 0$.

    \item \textbf{Stellar streams:} Disrupted low-mass galaxies which are torn and stretched along their orbits by  tidal forces from the host galaxy. 

    \item \textbf{Phase-mixed galaxies:} Disrupted low-mass galaxies which were once stellar streams, but are now merged with the host galaxy and occupy its equilibrium phase-space distribution.

    \item \textbf{Field galaxies:} Intact galaxies that are outside the virial radius (radii) of the isolated (paired) MW-like (MW/M31-like) host(s) at $z = 0$.
\end{enumerate}

We use three criteria based on the physical properties of streams to identify present-day stellar streams within $R_{\rm 200m}$ of the main host galaxy in each simulation. First, the number of star particles in the stream candidate must be larger than 100, which is equivalent to a stellar mass of $\sim$5 $\times$ 10$^5$ $M_\odot$. This is a practical limit on the required resolution to apply the second and third criteria, which are based on the application of Liouville's theorem, but is also a plausible lower limit on surface brightness for most streams. As we will discuss in Section \ref{subsec:det_stream_gal_results}, this selects streams with much lower stellar mass than required to create detectable tidal tails above our surface brightness cutoff (34 $\rm{mag}$ $\rm{arcsec^{-2}}$).

Stream stars are unbound from their progenitor and spread out in position as their local velocity dispersion decreases. Thus we require that the maximum pairwise separation between any two star particles in the stream candidate must be greater than 120 kpc, while its local velocity dispersion must be below a stellar-mass dependent threshold (see Equation 2 of \citealt{Panithanpaisal_2021}). If both the positions and velocities of the stars are widely distributed, we consider them to be phase-mixed, while if both are narrowly distributed then we conclude that the galaxy is still intact.  To summarize, we classify a galaxy as (a) a stream if it satisfies all three criteria; (b) a phase-mixed galaxy if it satisfies the first two criteria; and (c) a satellite galaxy if it satisfies just the first criterion. However, we manually reduce the stellar mass requirement to $10^{4.5}$ $M_\odot$ for the case of intact galaxies, since they are more compact in comparison to streams, and therefore are easier to detect. Additionally, for all our classified substructures within the virial radius, we exclude star particles associated with the present-day host galaxies themselves. We also exclude star particles associated with the present-day host galaxies themselves. For more details, we refer the reader to \citet{Panithanpaisal_2021, Kundu_2026}.

For the field galaxies, we select all the luminous halos with $M_\star \geq 10^{4.5}$ $M_\odot$ outside $R_{\rm 200m}$ of the main host galaxy within a region uncontaminated by low-resolution DM particles. This differs from the selection used in \citet{Kundu_2026}, where we imposed a distance cut of 2 Mpc from the center of each isolated MW-like galaxy, or from the geometric center of mass of the LG-like pairs. These criteria ensure that our sample does not suffer from numerical issues due to very low-mass halos and safely excludes any contaminated halos, while extending our field galaxy sample to larger distances. However, this leads to the field galaxies being drawn from a common distribution in case of the paired hosts, the consequences of which we discuss in Section \ref{sec:UVLFs}. Table \ref{tab:sim_data} shows the number of intact and tidally-disrupted galaxies in each of our 13 host galaxies in the simulations.

\begin{deluxetable*}{ccccccccc}
\tablecaption{\textbf{Properties of the FIRE-2 Simulations at $z = 0$.} }
\label{tab:sim_data}
\tablewidth{0pt}
\tablehead{
\colhead{Simulation Name} & \colhead{$M_{\rm 200m}$} & \colhead{$M_{\rm \star, 90}$} & \colhead{$R_{\rm 200m}$} & \colhead{$z_{\mathrm{MR_{3:1}}}$} &
\multicolumn{4}{c}{Number of}\\
\cline{6-9}
\colhead{} & \colhead{($\times 10^{12} \, M_\odot$)} & \colhead{($ \times 10^{10} \, M_\odot$)} & \colhead{($\rm{kpc}$)} & \colhead{} & 
\colhead{\parbox[c]{1.5cm}{\centering \vspace{0.5em} Satellite \\ Galaxies}} & \colhead{\parbox[c]{1.5cm}{\centering \vspace{0.5em} Stellar \\ Streams}} & \colhead{\parbox[c]{2cm}{\centering \vspace{0.5em} Phase-mixed \\ Galaxies}} & \colhead{\parbox[c]{1.5cm}{\centering \vspace{0.5em} Field \\ Galaxies}} \\
}
\startdata
m12i\tablenotemark{A} & 1.2 & 6.3 & 336 & 3.18 & 20 (5) & 13 & 17 & 7\\
m12b\tablenotemark{B} & 1.4 & 8.5 & 358 & 1.59 & 19 (3) & 13 & 10 & 18\\
m12f\tablenotemark{C} & 1.7 & 7.9 & 380 & 3.77 & 28 (4) & 12 &  29 & 26\\
m12c\tablenotemark{B} & 1.3 & 5.8 & 351 & 1.32 & 33 (9) & 11 & 7 & 55\\
m12m\tablenotemark{D} & 1.6 & 11.0 & 371 & 1.68 & 31 (10) & 13 & 6 & 40\\
m12r\tablenotemark{E} & 1.1 & 1.7 & 321 & 1.02 & 24 (5) & 5 & 6 & 20\\
m12w\tablenotemark{E}& 1.1 & 5.7 & 319 & 2.92 & 31 (12) & 5 & 14 & 21\\
\hline
Romeo\tablenotemark{B} & 1.3 & 6.6 & 341 & 5.70 & 23 (5) & 20 & 37 & \multirow{2}{*}{81}\\
Juliet \tablenotemark{B} & 1.1 & 3.8 & 321 & 4.72 & 22 (10) & 15 & 29 \\
\noalign{\vskip 5pt}
Romulus\tablenotemark{F}& 2.1 & 9.1 & 406 & 1.70 & 38 (7) & 14 & 25 & \multirow{2}{*}{58}\\
Remus\tablenotemark{F}& 1.2 & 4.6 & 339 & 0.98 & 20 (3) & 15 & 6 \\
\noalign{\vskip 5pt}
Thelma\tablenotemark{F} & 1.4 & 7.1 & 358 & 4.29 & 20 (8) & 11 & 24 & \multirow{2}{*}{62}\\
Louise\tablenotemark{F}& 1.1 & 2.6 & 333 & 3.32 & 27 (7) & 12 & 24 \\
\enddata
\tablecomments{\emph{Simulation Name}: References listed below; the top seven are isolated MW-like and bottom six (three pairs) are paired MW/M31-like. $M_{\rm 200m}$ and $R_{\rm 200m}$: total mass and spherical radius within which the mean density is 200 times the matter density of the Universe. $M_{\star, 90}$: stellar mass within a spherical radius that encloses 90\% of the stellar mass within 20 kpc. $z_{\mathrm{MR_{3:1}}}$: the redshift at which the stellar-mass ratio of the most massive halo to the second-most massive subhalo is 3:1, which is when the dominant host galaxy emerges. Numbers of surviving satellite galaxies ($M_\star \gtrsim 10^{4.5} M_\odot$; in parentheses, the number above the stream and phase-mixed mass cut ($M_\star \gtrsim 5 \times10^{5} M_\odot$)), stellar streams ($M_\star \gtrsim 5 \times 10^{5} M_\odot$), phase-mixed galaxies ($M_\star \gtrsim 5 \times 10^{5} M_\odot$), and field galaxies ($M_\star \gtrsim 10^{4.5} M_\odot$) are determined as described in Section \ref{subsec:classify_galaxies}.}
\tablerefs{(A) \citet{Wetzel_2016}, (B) \citet{Garrison-Kimmel_2019a}, (C) \citet{Garrison-Kimmel_2017}, (D) \citet{Hopkins_2018a}, (E) \citet{Samuel_2020}, (F) \citet{Garrison-Kimmel_2019b}.}
\end{deluxetable*}

\section{Creating Mock LSST Catalogs} \label{sec:mock_catalogs}

We use \textsf{py-ananke}\footnote{\url{https://py-ananke.readthedocs.io/en/main/}} \citep{Thob_2024}, a Python package based on the \textsf{ananke} framework \citep{Sanderson_2020}, to generate mock astrometric and photometric observations of synthetic stars in the satellites, streams, phase-mixed galaxies, and field galaxies identified in Section \ref{subsec:classify_galaxies}, as observed by LSST from within the host galaxy. We create a total of 13 mock catalogs: 7 from isolated MW halos and 6 from the three LG-scale simulations, each of which contains two MW-like halos. Here we provide a working summary of the steps employed to run the pipeline in this work and refer the interested reader to \citet{Sanderson_2020, Thob_2024} for more details. 

\subsection{Preparing the simulation data}
\label{subsec:prep_sim_data}

For each host galaxy, we establish a galactocentric coordinate system where the observer is placed on a solar viewpoint, specifically, a phase-space position for the local standard of rest (LSR). First, we determine the galactocentric coordinate frame, centered on and aligned with the stars in the galaxy such that the galaxy disk lies in the $X$-$Y$ plane and the disk angular momentum lies along the $Z$ axis. As discussed in \citet{Sanderson_2020}, we use the young stars (age $<$1 $\rm{Gyr}$) to define the disk angular momentum direction, since this selection tracks the plane of the thin gaseous disk but is more stable over time. The center position of the host galaxy is determined by applying the shrinking-spheres method to the star particles iteratively and the center-of-mass velocity is computed using the star particles within 15 kpc of the center of the host DM halo (identified by the halo catalog) \citep{Sanderson_2020}. 

In this frame we assume that the Sun is located at $R_{\odot} = 8.2$ $\rm{kpc}$ \citep{Bland-Hawthorn_2016}. Then, we compute the moment of inertia tensor using young star particles (age $<$1 $\rm{Gyr}$) inside the Solar radius. This establishes the $Z$ direction perpendicular to the plane of the galactic disk. As the norm for standard Galactocentric coordinates, we choose the orientation of the total angular momentum of the galactic disk to point in the $-Z$ direction, so that the host galaxy rotates clockwise. If the LSR position is on the $-X$ axis, the rotation of the galaxy carries it to the $+Y$ direction, consistent with the standard conventions used for Galactic coordinate systems. 

After establishing a galactocentric coordinate frame, we define a solar viewpoint and LSR in it. We initially start with a Solar position at $(-8.2, 0, 0)$ kpc and select all the star particles within 200 $\rm{pc}$ of this position. This allows us to define a sufficiently ``local'' LSR, with enough particles to avoid sampling noise. We then compute the median position and velocity using all the star particles within this ``local Solar neighborhood'' and use this to define our LSR position and velocity. This phase-space coordinate serves as our 6D observer position for LSST.

\subsection{Generating the density field}
\label{subsec:density_calculation}

\textsf{py-ananke} uses the local density of the selected star particles in phase space to determine the volume over which to distribute the mock stars sampled from that particle. We apply \texttt{EnLink-2.6} \citep{Sharma_2009} to the positions, velocities, and masses of the selected star particles to compute the phase-space density at the location of each simulated star particle\footnote{We use the following specific flags while computing the densities: \texttt{-dsbcm --dim=6 --ngb=16 --dsuffix=\_d6n16} \\ \texttt{d}: calculate density, \texttt{s}: use entropy-based estimation, \texttt{b}: use bootstrapping, \texttt{c}: ``cubic'' densities, \texttt{ngb}: number of nearest neighbors, \texttt{dsuffix}: informational suffix to the file}. The six-dimensional densities are estimated using the adaptive, entropy-based density estimation scheme described in \citet{Sharma_2009} with 16 nearest neighbors, and smoothed using bootstrapping. The estimated densities are converted into equivalent single-dimensional, ``cubic'' densities in the position and velocity subspaces and then inverted to return a local estimate for the bandwidth of a spherical Epanechnikov kernel distribution in position and velocity, such that sampling this distribution reconstructs the input density distribution. The resulting kernel lengths are written to a file which we use as inputs to \textsf{py-ananke}. Because the density estimate uses 16 nearest neighbors, we restrict the sample to substructures with more than 20 star particles. This only removes some of the intact low-mass galaxies with stellar masses between $10^{4.5}\,M_\odot$ and $10^{5}\,M_\odot$ and not the streams, phase-mixed galaxies and higher-mass intact galaxies. We classify these low-mass galaxies as undetectable, but keep them in the simulated population.


\subsection{Creating a mock LSST catalog using py-ananke} 
\label{subsec:run_py-ananke}

The galactocentric positions and velocities and the list of kernel lengths for the selected star particles are passed to \textsf{py-ananke} along with their masses, ages, and iron abundances [Fe/H], which are used to determine the number of stars spawned from the single-age, single-metallicity stellar population represented by each particle. We also pass additional information from the parent star particle on to each synthetic star: chemical abundances for He, C, N, O, Ne, Mg, Si, S and Ca, $\alpha$ abundances (mean abundance using O, Mg, Si, Ca), and the distance of the star particle from the host-galaxy center at formation (also known as the birth radius).

\textsf{py-ananke} samples stellar masses from a Kroupa IMF \citep{Kroupa_2001} and calculates their absolute magnitudes in the LSST filters using PARSEC isochrone models \citep{Bressan_2012} matched to the metallicity and age of each star particle. Specifically, we use a grid of isochrones for non-rotating single stars generated using PARSEC version 1.2S \citep{Chen_2014, Tang_2014, Chen_2015}, incorporating COLIBRI models of asymptotic giant branch evolution \citep{Marigo_2013, Rosenfield_2016, Pastorelli_2019, Pastorelli_2020} with $\eta_{\rm{inTPC}} = 10$ \citep{Marigo_2017} and $\eta_{\rm{Reimers}}=0.2$. Circumstellar dust is modeled using scaling relations from \citet{Marigo_2008} for the recommended multi-component dust compositions: 60\% silicate/40\% AlOx for M stars and 85\% AMC/15\% SiC for C stars \citep{Groenewegen_2006}. The isochrone grid spans 71 age values evenly spaced from $\log_{10}(\rm{age}/\rm{yr}) = 6.6$ (3.98\,Myr) to $\log_{10}(\rm{age}/\rm{yr}) = 10.1$ (12.6\,Gyr) and 34 metallicities from $Z = 0.0001$ to $Z = 0.03$ (approximately equivalent to $-2.3 < \mathrm{[M/H]} < 0.2$). The age and metallicity of each star particle are mapped to the closest isochrone in the grid for sampling; in other words, we do not interpolate between isochrones. For the few star particles that fall outside of the range spanned by the model isochrones, we use the nearest edge of the grid.

We also define our survey parameters for \textsf{py-ananke}, based on projections for LSST. We set the \texttt{observer} parameter to the location (in kpc) of the observer in galactocentric coordinates as calculated in Section \ref{subsec:prep_sim_data}, which roughly corresponds to $(-8.2, 0, 0)$ kpc. 
We set \texttt{fsample=1} for a full-sized survey. We choose \texttt{rshell}=4000 kpc to sample stars from star particles in the tidal arms of the streams that extend beyond the host halo, as well as the field galaxies. 
We calculate magnitudes of the sampled stars in the six bands of the LSST filter system (\texttt{photo\_sys = `LSST'}) and specify the limiting magnitude of the survey in one LSST filter with the \texttt{cmd\_magnames} and \texttt{cmd\_box} parameters. More details about these parameters are available at --- \url{https://py-ananke.readthedocs.io/en/main/}

Since LSST has different limiting magnitudes in its different filters, which stars are selected after sampling from the IMF depends on the filter and magnitude limit we choose. We choose to accept stars into the survey that are detected in at least one of the six bands; since the red filters are most efficient, we set the limiting magnitude in the $z$ band to 29 for the preliminary survey, before making the final cut. This is somewhat fainter than the published 10-year co-added apparent magnitude limit for LSST in this band (25.6; \citealt{Bianco_2022}) in order to allow for the spread in distances of the stars spawned by each star particle and the stochastic rounding used during the sampling of the IMF to generate an integer number of stars from each particle. We achieve significant speedup by estimating the lower limit in mass that corresponds to the faint limit of the survey at the location of each star particle and limiting the IMF sampling to this range, but for galaxies with very few simulated star particles this simplification can combine with the distance spread and stochastic sampling to erroneously produce zero stars above the strict limiting magnitude of the survey. Choosing an apparent magnitude which is a bit fainter than our survey limit ensures that we get some synthetic stars for these very low-mass galaxies. Before analyzing our mock catalogs, we then re-select stars with apparent magnitudes brighter than the LSST faint limit in at least one band. Hence, to generate a mock catalog using \textsf{py-ananke} it is advisable to use the most sensitive filter for defining the magnitude limit, and to choose a preliminary apparent magnitude limit a bit fainter than the desired final faint limit in that filter.
 
Running \textsf{py-ananke} using the above methods generates mock LSST catalogs for the substructures identified in each of the 13 simulated MW analogs in Table \ref{tab:sim_data}, assuming perfect observation of their stellar populations without any foreground stars. For a few very low-mass galaxies, we get zero synthetic stars based on our survey parameters; we assume these to be undetectable. The data contained in the mock catalogs is described in detail in Appendix \ref{appendix:mock_data}. In brief, we generate sky positions, parallaxes, proper motions, and radial velocities; absolute and apparent magnitudes in the six LSST bands; stellar parameters and chemical abundances; and indices mapping each star to its parent star particle in the simulation.

To standardize file size and for greater accessibility, the resulting catalogs are saved in a series of files corresponding to a multiscale Healpix tessellation of the sky in Galactic coordinates. The tessellation follows the Multi-Order Coverage (MOC) mapping convention defined by \citet{Fernique_2022} where a Healpix NUNIQ indexing scheme allows for each Healpix cell to be uniquely identified across all resolution levels. The determination of the tessellation is done iteratively through the resolution levels using the distribution of star particles from the simulation and their initial masses that were used as input of \textsf{py-ananke}. Starting at level 0 with the full set of particles, the process checks all Healpix cells of that level (12 of them at level 0) and compares for each of them the total mass from the particles they contain to a threshold mass of $10^8 \, M_\odot$: if the total is below the threshold, the cell's NUNIQ id is saved and the corresponding particles are removed from the set of particles to consider in the next iteration. Once all cells of the level have been considered, if particles still remain in the set, the process iterates the same procedure in the next resolution level with the remaining particles, and continues to do so in the deeper levels until no particles are left. The result is a set of NUNIQ ids corresponding to the full multiscale Healpix tessellation that exhaustively covers the entire sky. While this process does not allow for the consideration of particles kernels overlapping over multiple cells where their center are sometimes not located --- considering this would require a more involved algorithm --- it provides a good first approximation of a multiscale tessellation that matches the expected distribution of synthetic stars for accessibility purposes. 

The mock catalogs generated in this work contain perfect information for all the synthetic stars that would be observed by LSST, but do not account for either galactic extinction and reddening or LSST's photometric errors. We do not expect extinction and reddening to have a large effect on our results since most of the streams are well above or below the galactic disk. In the future, we plan to use LSST's \texttt{StreamObs}\footnote{\url{https://github.com/LSSTDESC/streamobs}} software to model LSST uncertainties. Although our mock catalogs adopt the 10-year co-added magnitude cuts for completeness, we use the 1-year magnitude cuts for making predictions in this work since the proper motions of the stars will likely play a significant role in assigning them to streams.

\subsection{Data availability}

All our LSST synthetic observations of satellite galaxies, stellar streams, phase-mixed galaxies, and field galaxies, are available via Globus\footnote{\globuslink}. 
Anyone can access the data after logging in with a free Globus account. We include the substructures from all the 7 isolated MW-like hosts (simulation names listed as m12i, m12b, m12f, m12c, m12m, m12r, and m12w), and the 3 MW/LG-like hosts (simulation names listed as RJ, RR, TL). Within each host directory we have the synthetic observation of a classified substructure, with its own subdirectory in the format: \texttt{<simname>\_<cflag>\_<ID>\_Rubin\_lsr\_axes}. In case of the isolated suite, \texttt{simname} corresponds to the simulation name, e.g., \texttt{m12i}. In case of the paired suite it also corresponds to the simulation name but with an additional flag of the host, e.g., \texttt{RJ\_host1} or \texttt{RJ\_host2}, identifying the respective host. \texttt{cflag} can be either \texttt{cat} or \texttt{hal}, where \texttt{cat} includes the stellar streams, phase-mixed galaxies, and satellite galaxies with $M_\star \gtrsim 5 \times 10^5 \, M_\odot$, and \texttt{hal} includes the field galaxies and the lower-mass satellites galaxies ($10^5 \, M_\odot \lesssim M_\star \lesssim 5 \times 10^5 \, M_\odot$). We include the \texttt{ID} of each identified substructure in the \texttt{<simname>\_object\_ids.pkl} in each simulation directory. The key names \texttt{stream}, \texttt{sat}, and \texttt{phase\_mixed} belong to the category \texttt{cat}, while \texttt{other\_dwarf\_indices} and \texttt{field\_gal\_indices} belong to the \texttt{hal} category and corresponds to the halo indices from the halo catalogs. The key names \texttt{very\_low\_mass\_other\_dwarf\_indices} and \texttt{very\_low\_mass\_field\_dwarf\_indices} includes the low-mass halos with $M_\star \gtrsim 10^{4.5}\, M_\odot$ but have less than 20 star particles. These halos do not have any synthetic observation due to our selection criterion of star particles. Each synthetic observation directory includes: the hdf5 files \texttt{\seqsplit{survey.<simname>\_<cflag>\_<ID>\_Rubin\_lsr\_axes.N.h5}}, where \texttt{N} is the Healpix partition ID; the survey parameter file \texttt{survey\_params}; and the density file in ebf format. 

\section{Detectability of Streams and Intact Galaxies} 
\label{sec:detectability}

We follow an approach similar to that of \citet{Shipp_2023} to estimate the detectability of the mock streams and satellites, with a few important differences. \citet{Shipp_2023} estimated the detectability of the streams and satellites for DES using the respective substructure classifications from \citet{Panithanpaisal_2021}. In this work, we use the classified substructures from \citet{Kundu_2026}, which contain some additional objects which are massive ($M_\star \gtrsim 10^9 \, \rm{M_\odot}$) and morphologically complex, and also the low-mass satellites ($10^{4.5} \, \rm{M_\odot} \lesssim M_\star \lesssim 10^{6} \, \rm{M_\odot}$) --- both of which are excluded from \citet{Panithanpaisal_2021}. Additionally, we also estimate the detectability of field galaxies based on the criteria discussed in \citet{Mutlu-Pakdil_2021} for LSST-based searches of low-mass galaxies beyond the MW and LG. We do not estimate the detectability of the phase-mixed galaxies and defer that to future work.

\subsection{Stream Detectability} \label{subsec:stream_detectability}

As described in \citet{Shipp_2023}, we determine the detectability of the streams following a method motivated by matched-filter stream searches in the DES data \citep{Shipp_2018}. The detectability of the streams can be estimated using the average surface brightness from the observed stars within the survey magnitude limits. Streams that have low luminosity and are more diffuse have fainter surface brightness, making it more difficult to disentangle them from background and foreground stellar populations. This method also takes into account that more distant streams will have fewer detectable stars, and hence will have a fainter observed surface brightness. Instead of using a converted \textit{V}-band magnitude to compute the average surface brightness of the streams as \citet{Shipp_2023} did for DES, we directly use the LSST \textit{g}-band, which is close to the \textit{V}-band. Since it is difficult to estimate the surface brightness limit for resolved stellar populations in LSST before the availability of real survey data, we keep the surface brightness limit as $\mu_{g} \leq 34 \, \rm{mag \,\, arcsec^{-2}}$, the same as the surface brightness limit in $V$-band used in \citet{Shipp_2023}. We restrict our results to the magnitude limits for 1-year exposure and hence obtain conservative estimates with respect to the survey depth. Data from additional years and improved detection methods will reach deeper magnitudes and fainter surface brightnesses. Users can explore this avenue, as our mock catalogs go deeper in magnitudes than the 1-year limits.

\begin{figure*}[ht!]
    \centering
    \includegraphics[width=\linewidth]{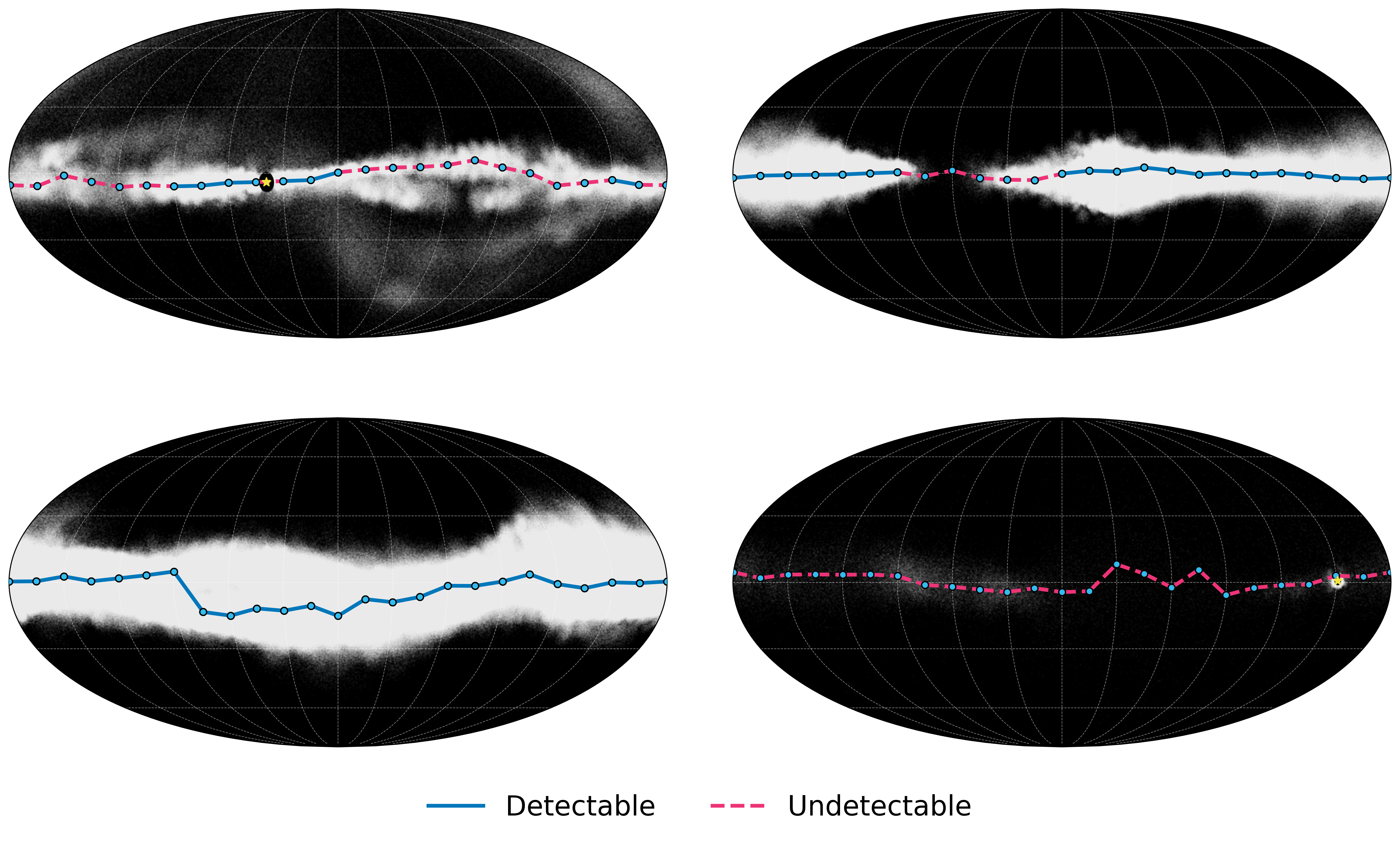}
    \caption{Examples of mock LSST stellar streams showing the number density of stars. Each panel shows the mock observation of one stream projected along the stream coordinates $(\phi_1, \phi_2)$. We divide each stream into segments of 15 deg, and calculate the average surface brightness within the 1$\sigma$ Gaussian width to estimate the detectability. Blue segments are detectable ($\mu_{g} \leq 34 \, \rm{mag \,\, arcsec^{-2}}$), and magenta dashed segments are undetectable. Any surviving bound progenitor is masked within 4$r_{\rm{1/2} }$ (black patches in top left and bottom right) and its position is shown by a yellow star.
    {\it Top left}: a stream with a bound progenitor and detectable tidal tails. {\it Top right}: a stream without any bound remnant and a significant fraction of detectable tidal tails. {\it Bottom left}: a stream in which all segments are detectable along the tidal tails. {\it Bottom right}: a stream with a detectable bound remnant (visible in the unmasked version which is not shown here) but no detectable tidal tails. This would be detectable as a satellite.}
    \label{fig:streams_examples}
\end{figure*}

In order to determine the average surface brightness of the tidal tails of a mock FIRE stream, we start by masking all the stars within $4r_{\rm{1/2}}$ of the bound progenitor (if any), using the $r_{\rm{1/2}}$ parameter derived in Section \ref{subsec:intact_gal_detectability}. Then we follow the exact approach of \citet{Shipp_2023} and convert to a coordinate system in which the stream lies approximately along the equatorial plane, where $\phi_1$ is the coordinate along the stream and $\phi_2$ is the coordinate perpendicular to the stream. The stars belonging to a long, convoluted stream may not share the exact same orbital plane. Hence, we estimate the best-fit RA and DEC of the orbital pole ($\alpha$, $\delta$) for each stream by minimizing the quantity $\mathcal{D}(\alpha, \delta) = \sum_i \left|\frac{\pi}{2} - \mathcal{D}_i(\alpha, \delta) \right|$, where $\mathcal{D}_i(\alpha, \delta)$ is the angular distance from the pole to each stream member star.

To capture the variation in surface brightness along the stream, we divide it into segments of equal length (15 deg) and compute the detectability of each one individually. Then, to compute the $\phi_2$ position of each endpoint, we take stars that are within 3 deg of $\phi_1^{\rm{end}}$. If this region contains fewer than 50 stars, we exclude this endpoint as it is likely empty, and automatically flag its neighboring segments as undetectable. For sufficiently populated regions, we identify the corresponding $\phi_2^{\rm{end}}$ to be the peak in the $\phi_2$ distribution of the selected stars. After determining the detectability of each of these segments individually, we classify a stream as detectable provided it has at least one detectable segment (after masking the bound progenitor). Any system with more than one detectable segment is considered to be a single detectable stream --- consistent with how disconnected but physically associated segments of known Milky Way dwarf galaxy streams are now treated as one single stream. Examples include ATLAS-Aliqa Uma \citep{Koposov_2014, Shipp_2018, Li_2021}, Orphan-Chenab \citep{Grillmair_2006, Belokurov_2007, Shipp_2018, Koposov_2019}, and Sagittarius \citep{Ibata_1994, Majewski_2003, Hasselquist_2019, Johnson_2020}.

Within each segment, we apply another coordinate transformation such that the great circle joining the two endpoints aligns with the equatorial plane. Then, we measure the segment width in $\phi_2$ by fitting the peak interval --- defined as the smallest $\pm 1 \sigma$ interval enclosing the peak of the distribution. For our analysis, we restrict to only the inner $\pm 1 \sigma$ range to avoid bias in the detectability estimate from the diffuse components at large $\phi_2$.

Unlike \citet{Shipp_2023}, who converted DES \textit{g}- and \textit{r}- band magnitudes to the \textit{V} band, we estimate the total luminosity directly in the LSST \textit{g}-band (absolute magnitude $M_g$). Using the total luminosity and area of each segment, we calculate the average surface brightness ($\mu_g$). We use a detectability limit of $\mu_{g} \leq 34 \, \rm{mag \,\, arcsec^{-2}}$. In Figure \ref{fig:streams_examples}, we show some examples of mock streams projected along the stream coordinates. The top left panel shows a stream with both a detectable bound progenitor and detectable tidal tails and the top right panel shows a stream with no bound remnant but with most of the tidal tails being detectable. The bottom left panel shows a stream which is entirely detectable, and the bottom right panel shows a stream which only has a detectable bound part without any tidal tails --- meaning it will be detectable as a satellite.

We emphasize that our detectability results may change depending on the actual surface brightness limits LSST will achieve. Additionally, we do not consider background galaxies to estimate detectability. This requires star-galaxy separation, particularly at faint magnitudes, where unresolved galaxies can contaminate the foreground stellar populations. Hence, our estimates are optimistic by neglecting background contamination, and users of our catalog are encouraged to explore to what extent the results can change when unresolved background galaxies are included.

\subsection{Intact Galaxy Detectability} \label{subsec:intact_gal_detectability}

We also estimate the detectability of intact galaxies and stream progenitors in our catalog at $z = 0$. This category includes the satellites identified by our pipeline ($M_\star \gtrsim 5 \times10^5 \,\, \rm{M_\odot}$; Section~\ref{subsec:classify_galaxies}), the additional low-mass satellites ($10^{4.5} \, \rm{M_\odot} \lesssim M_\star \lesssim 5 \times10^5 \,\, \rm{M_\odot}$) and the field galaxies identified from the \textsc{Rockstar} halo catalogs. Some of the streams at $z = 0$ still have bound parts, which are identified as bound halos in the halo catalogs. We call them ``stream progenitors'' and also estimate their detectability. If the stream has no detectable tidal tail using the method described in Section~\ref{subsec:stream_detectability}, but has a detectable bound part, we classify it as a satellite in our subsequent analysis. 

Using the mock LSST observations of these objects, we follow a similar approach to \citet{Bechtol_2015, Drlica-Wagner_2015, Drlica-Wagner_2020, Shipp_2023} and fit an ellipsoidal Plummer \citeyearpar{Plummer_1911} profile to measure the half-light radius ($r_{1/2}$). However, unlike \citet{Shipp_2023}, we do not use the code \textsf{ugali}\footnote{\url{https://github.com/DarkEnergySurvey/ugali}} to compute the parameters. We instead fit an ellipsoidal Plummer profile to the spatial distribution of detected stars using Nelder-Mead optimization \citep{Nelder-Mead} to get $r_{1/2}$ in the V-band, for which we convert the $g$- and $i$-band LSST magnitudes to visual magnitudes using the relations in \citet{LSST_conversions}. Furthermore, we calculate the total luminosity ($M_{\rm V}$) by taking the median distance of the mock member stars as the distance to the intact galaxy. Unlike \citet{Shipp_2023}, we do not correct for the contribution to the luminosity from unobserved stars (due to LSST magnitude limits) using the isochrones in \textsf{ugali}.  We will show in Section \ref{subsec:size_lum_relation} that despite these approximations and simplifications, our results still agree well with the observed properties of MW satellites above our simulation resolution limits. However, we encourage the users of our mock catalogs to try more sophisticated methods for calculating these properties. 

The detectability of satellites can be estimated using an analytical approximation as a function of absolute magnitude, half-light radius, and distance. This was presented by \citet{Drlica-Wagner_2020} for DES and then updated for LSST in \citet{Tsiane_2025}. We use the derived parameters above to estimate the detectability of satellites and stream progenitors within the virial radius of the host galaxy. Combining these results with those for detectable streams (Section \ref{subsec:stream_detectability}), we can divide our simulated streams into streams with detectable tidal tails, detectable progenitors only (streams mistakenly classified as satellites), or no detectable component.

For the field galaxies we follow a different approach. Using our estimates of the absolute magnitude and half-light radius, we compare with the size-luminosity limits presented in \citet{Mutlu-Pakdil_2021} for dwarf galaxies out to $\sim$3.5\,Mpc using LSST. We use their 50\% detectability regions in $r_{1/2}$--$M_V$ in the two closest distance bins ($<1.5$\,Mpc, $<3.5$\,Mpc) and flag the galaxies within these regions as detectable. We highlight that their detectability estimates are based on 10 years of LSST data, while our properties are calculated using 1-year magnitude cuts. This means our detectability estimates of field galaxies are conservative and will only improve if using the 10-year magnitudes.

\begin{figure*}[ht!]
    \centering
    \includegraphics[width=\linewidth]{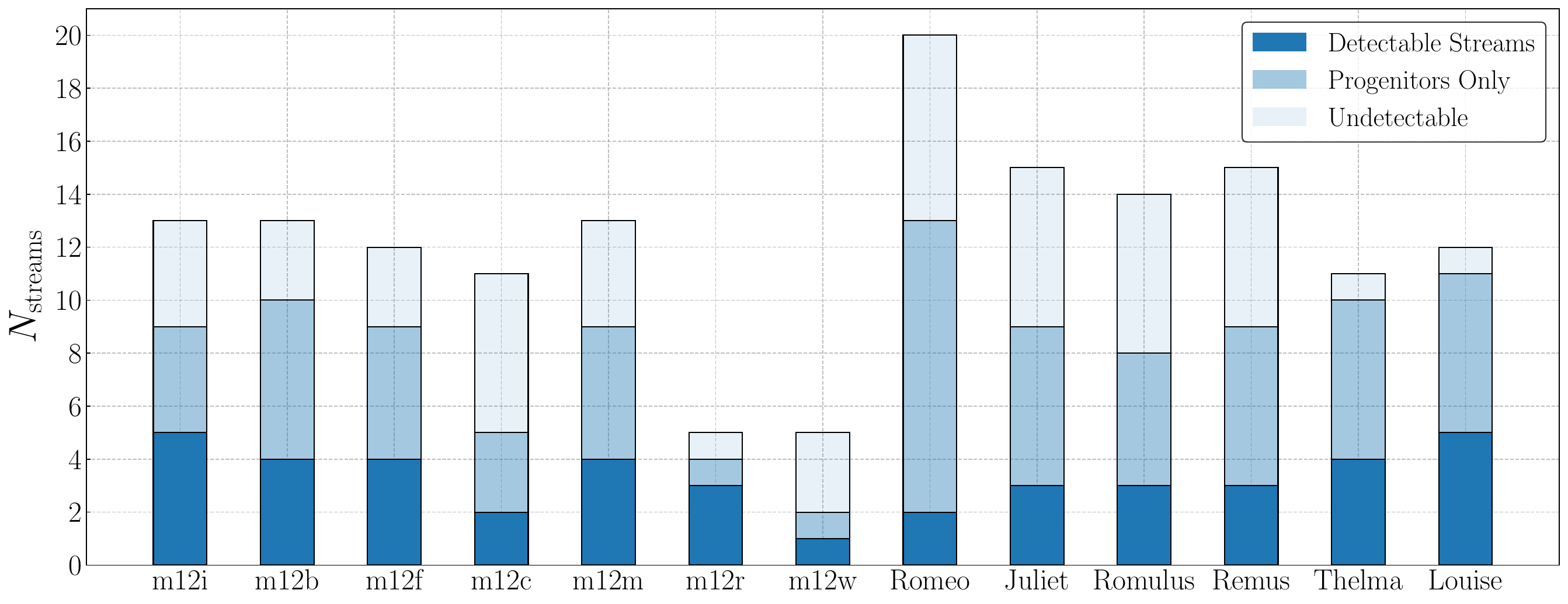}
    \caption{Total and detectable number of streams in each of the 13 FIRE host galaxies. The bottom segment represents the number of streams with detectable tidal tails, the middle segment shows the number of streams with only detectable bound progenitors (i.e., as satellites), and the top segment represents the rest of the streams which are undetectable. The total height of each bar shows the total number of streams in each host galaxy.}
    \label{fig:number_of_streams}
\end{figure*}

\section{Present-day Properties of Streams and Intact Galaxies} 
\label{sec:present-day_props}

\subsection{Detectability Estimates of Streams and Intact Galaxies}
\label{subsec:det_stream_gal_results}

Using our detectability estimates for the stream populations and intact galaxies, we compute the number of detectable systems. In Figure \ref{fig:number_of_streams}, we show the number of detectable streams in each of the 13 host galaxies. The bottom segment of each bar shows the number of streams with detectable tidal tails, the middle segment represents the number of streams with only detectable progenitors (i.e., would be detected as satellites), and the top segment shows the number of undetectable streams (i.e., neither the tails nor the bound progenitor is detectable). The total height of each bar shows the total number of streams with $M_\star \gtrsim 5 \times 10^5\,M_\odot$ identified in each host galaxy. The median number of streams with detectable tidal tails is 3, which is consistent with the predictions of \citet{Shipp_2023}. Across the 13 simulations, we find an average of $\approx29\%$ of the streams to be detectable with tidal tails. We find that many of the streams in this stellar mass range and at this luminosity would remain undetectable even with LSST (within our chosen surface brightness limit). However, we would also mistake many of the streams for intact satellites, since their tidal tails are too diffuse to detect. We find on average $\approx38\%$ of the streams across the 13 hosts to be detectable as intact satellites. This can have important implications when reconstructing the high-$z$ UVLFs using the near-far technique. If the bound component contains only a small fraction of the total stellar mass of the stream, the inferred UV magnitude of its progenitor at high-$z$ can be biased unless the tidal tails are properly accounted for. We briefly explore the impact on the SFHs in Section~\ref{sec:SFHs}, but defer a detailed analysis to future work. We expect the percentages of streams with detectable tidal tails and streams detectable as satellites to change when using a higher-resolution simulation (within the same stellar-mass limit) as some of the systems classified into one category in the lower-resolution simulation can be re-classified into a different category in the higher-resolution simulation. For example, \citet{Meziani_2026} find two more stellar streams above the same mass resolution in the higher-resolution m12i simulation than the lower-resolution simulation used here, which would change our detectability estimates somewhat. These two streams are likely labeled as phase-mixed objects by our pipeline in the lower resolution simulations we study here. On the other hand, we expect the resolution to have minimal effect on gravitationally bound objects (satellite galaxies and the parts of stream progenitors that remain bound). Thus, while the number of detected streams we predict may be a lower limit, the overall contribution of partially or fully disrupted galaxies remains significant, even if a few of them are classified differently at higher resolution.

\begin{deluxetable}{lcc}
\tablecaption{\textbf{Detectable fractions across the 13 FIRE host galaxies.}}
\label{tab:det_fractions}
\tablewidth{0pt}
\tablehead{
\colhead{Substructures} & \colhead{$M_\star \gtrsim 10^{4.5}\,M_\odot$} & \colhead{$M_\star \gtrsim 10^{5}\,M_\odot$}
}
\startdata
Satellites & 59\% & 96\% \\
Bound stream progenitors & 89\% & 95\% \\
Field galaxies ($<1.5$\,Mpc) & 47\% & 71\% \\
Field galaxies (1.5--3.5\,Mpc) & 28\% & 46\% \\
Field galaxies ($<3.5$\,Mpc) & 42\% & 64\% \\
\cutinhead{Streams ($M_\star \gtrsim 5\times10^5\,M_\odot$)}
With detectable tidal tails & \multicolumn{2}{c}{29\%} \\
Detectable only as satellites & \multicolumn{2}{c}{38\%} \\
\enddata
\tablecomments{Fractions are averages across the 13 hosts (10 for field galaxies; see text), estimated using 1-year LSST magnitude limits and $\mu_g \leq 34$\,mag\,arcsec$^{-2}$ for tidal tails (Section~\ref{sec:detectability}).}
\end{deluxetable}

We next consider the intact galaxies and the bound components of streams. Across the 13 hosts, an average of $\approx59\%$ of the satellites with $M_\star \gtrsim 10^{4.5}\,M_\odot$ are detectable. The undetectable satellites are predominantly low-mass systems close to our resolution limit, including those with fewer than 20 star particles for which we do not generate mock stars (Section~\ref{subsec:density_calculation}). Increasing the stellar mass cutoff to $M_\star \gtrsim 10^{5}\,M_\odot$ improves the detectability fraction to $\approx$96\%, indicating that the undetectable satellites are almost entirely confined to $10^{4.5}\,M_\odot \lesssim M_\star \lesssim 10^{5}\,M_\odot$ and are therefore limited primarily by simulation resolution rather than by LSST sensitivity. For the streams that have a surviving bound progenitor at $z = 0$, $\approx$89\% of the progenitors with $M_\star \gtrsim 10^{4.5}\,M_\odot$ are detectable. The fraction increases to $\approx$95\% when we consider the progenitors with $M_\star \gtrsim 10^{5}\,M_\odot$. For the field galaxies, we average over the 7 isolated halos and the 3 LG-like pairs. In case of the paired suite, we consider the field galaxies to be outside the virial radii of the two main hosts, but use the properties as estimated from the more massive host to avoid overcounting. This is similar to the observed local field dwarfs --- observing the field galaxies of both MW and M31 from MW. We find that $\approx42\%$ of the field galaxies with $M_\star \gtrsim 10^{4.5}\,M_\odot$ out to $\sim$3.5\,Mpc are detectable, decreasing from $\approx$47\% within 1.5\,Mpc to $\approx$28\% between 1.5 and 3.5\,Mpc. For $M_\star \gtrsim 10^{5}\,M_\odot$, these fractions increase to $\approx64\%$, $\approx71\%$, and $\approx46\%$, respectively. These fractions are summarized in Table~\ref{tab:det_fractions}.

\begin{figure*}[ht!]
    \centering
    \includegraphics[width=\linewidth]{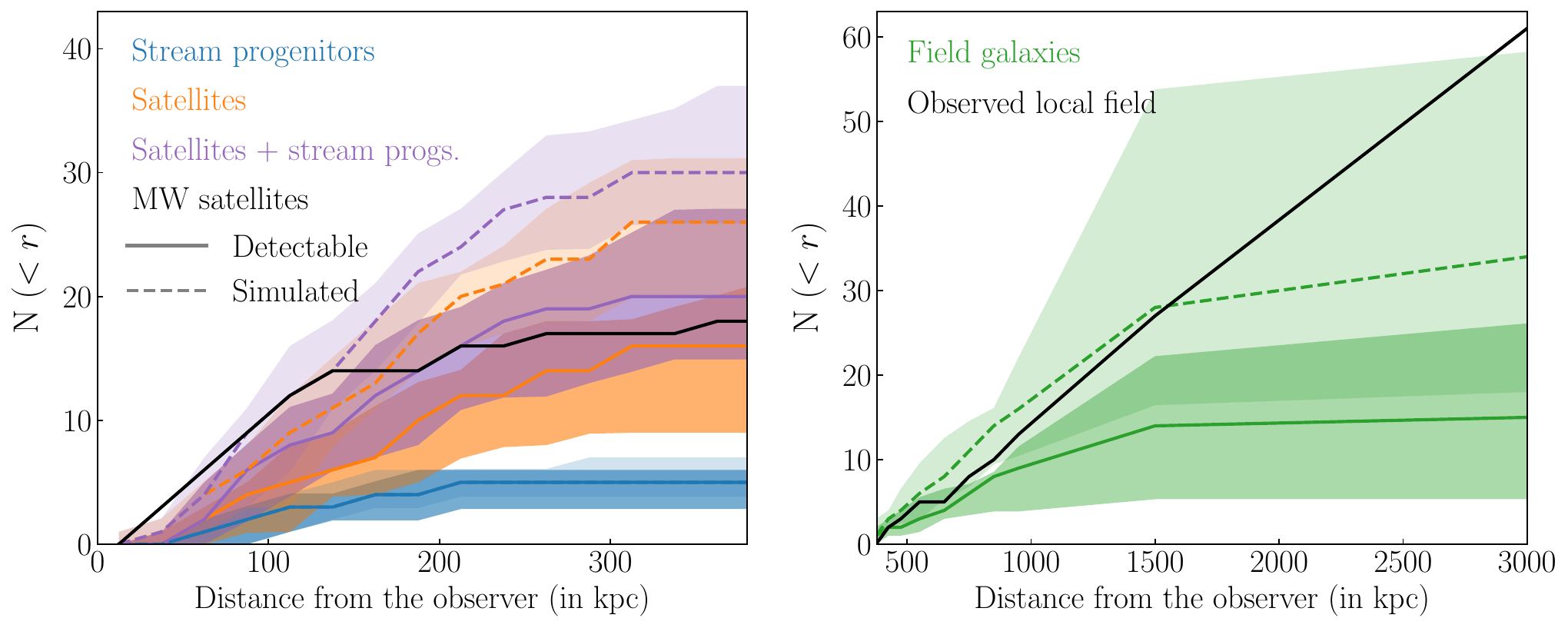}
    \caption{The cumulative number of intact galaxies and stream progenitors as a function of distance from the observer. The solid [dashed] lines with darker [lighter] shaded regions represent the median number and 68\% halo-to-halo scatter of detectable [simulated] populations, across the 13 host galaxies (10 in case of field galaxies common to both hosts in the paired suites). \textit{Left:} The number of satellites (orange), stream progenitors (blue) and their combined population (purple), compared with the observed MW satellites (black) from LVDB. \textit{Right:} The number of field galaxies (green) along with the local field dwarfs taken from LVDB. The radial distribution of combined population of detectable satellites and stream progenitors agrees well with that of the observed MW satellites within the halo-to-halo scatter. The distribution of detectable field galaxies also agrees well with that of the observed local field dwarfs up to $\sim800$\,kpc. See text for details.}
    \label{fig:radial_profile}
\end{figure*}

In Figure~\ref{fig:radial_profile}, we show the cumulative number of detectable galaxies as a function of distance from the observer and compare them against that of the MW satellites (left panel) and local field ($<3$\,Mpc) galaxies (right panel) taken from the Local Volume Database (LVDB)\footnote{\url{https://github.com/apace7/local_volume_database}} \citep{Pace_2025}. The lines show the median number of satellites (orange), stream progenitors (blue), satellites + stream progenitors (purple), and field galaxies (green), with the shaded regions showing the 68\% halo-to-halo scatter across the 13 host galaxies (10 for field galaxies). The solid lines and the darker shaded regions represent the detectable population, while the dashed lines and the lighter shaded regions represent the simulated population. We also show the observed distribution of MW satellites and the local field dwarfs (black) with $M_\star \geq 10^{4.5}\,M_\odot$ taken from LVDB. In the left panel, the observed MW satellite distribution is consistently above the median distribution of the detectable population and is within the 68\% halo-to-halo scatter only beyond $\sim$220\,kpc. However, it agrees well with the distribution of the combined population of detectable satellites and stream progenitors (streams detectable as satellites) within the halo-to-halo scatter beyond $\sim$150\,kpc, indicating that some of the observed MW satellites could be bound remnants of disrupted systems that would still appear as satellites even with LSST. The small discrepancy within $\sim$150\,kpc is mainly due to our sample selection for generating the mock catalog. As discussed in Section~\ref{subsec:density_calculation}, we generate the mock catalogs for substructures with more than 20 star particles. Thus, the simulated low-mass intact galaxies and the bound stream progenitors with stellar masses within $10^{4.5}\,M_\odot \lesssim M_\star \lesssim 10^{5}\,M_\odot$, but with fewer than 20 particles, are undetectable. This biases the distribution from the detectable population low, while the simulated distribution agrees well with the observed MW satellite distribution within 150\,kpc. We further confirm this by selecting the low-mass intact galaxies, stream progenitors, and the MW satellites from LVDB with $M_\star \geq 10^5\,M_\odot$. Figure~\ref{appendix_fig:radial_profile} shows that the observed distribution agrees well within the 68\% halo-to-halo scatter of the combined population of detectable satellites and stream progenitors over the entire distance range. In this case, the simulated distribution is very similar to the detectable distribution across all three populations.

\begin{figure*}[ht!]
    \centering
    \includegraphics[width=\linewidth]{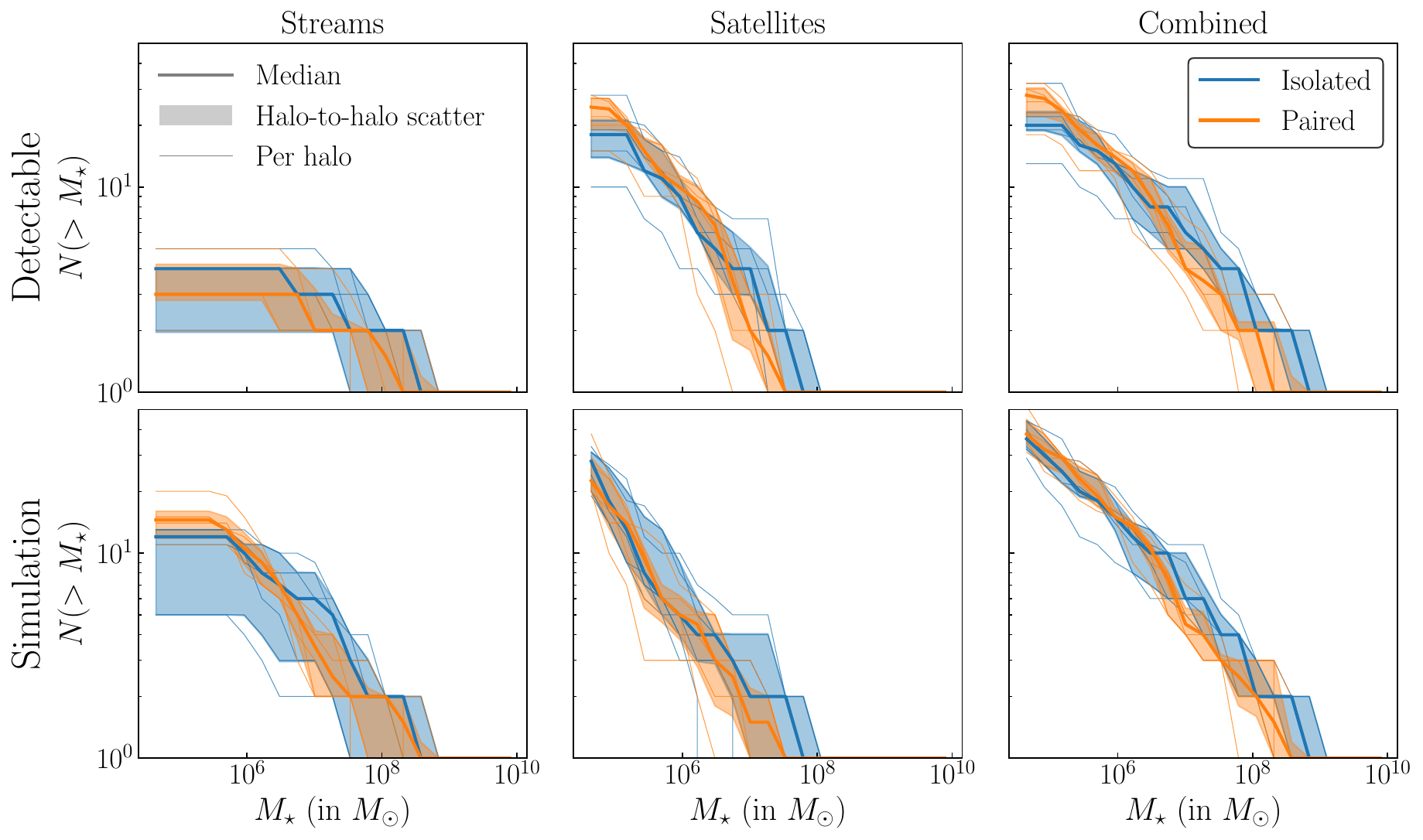}
    \caption{Present-day cumulative stellar mass functions (SMFs) for the streams and satellites in the isolated (blue) and paired (orange) suites, constructed using only the detectable substructures (top) compared with the total SMFs from the simulations (bottom). The left column shows only the streams, the middle column shows only the satellites (in the top row, this includes the streams without detectable tidal tails), and the right column shows the combined population of streams and satellites. The thin lines show the individual SMFs from each host galaxy, with the thick lines and the shaded regions showing their median and 68\% halo-to-halo scatter.}
    \label{fig:z0_SMFs}
\end{figure*}

We find these results to be consistent with simulated radial distributions in \citet{Samuel_2020}, but also note some discrepancies (see their Figure 3). We note that our median curve is below the observed MW satellite distribution within 150\,kpc, while they have a better agreement. This is because our observed MW satellite population from LVDB with $M_\star \geq 10^5\,M_\odot$ is not the same as theirs, since LVDB includes more recently discovered satellites which is missing in their analysis. Furthermore, their median radial distributions are combined across the last 1.3 Gyr in the simulations, while we only show them at the present day. Additionally, the underprediction from the detectable/simulated population within 150\,kpc could be physical. The presence of the LMC in the MW makes its radial distribution more concentrated within 150\,kpc at the present day than the typical simulated profiles \citep{Patel_2024}. Since not all of the FIRE host galaxies have an LMC-like counterpart, the median curve underpredicts the observed distribution. A few host galaxies which have an LMC-like satellite agree well with the observed distribution (see the thin lines in Figure~\ref{appendix_fig:radial_profile}).

In the right panel of Figure~\ref{fig:radial_profile}, the radial distribution of detectable field galaxies underpredicts the observed local field distribution beyond $\sim$800\,kpc. The simulated distribution agrees well, indicating most of our field galaxies at these distances would go undetected. Our zoom-in simulations, particularly the isolated suite, do not capture the entire local field population within the resolution limit. Since zoom-in simulations are run at a higher-resolution after selecting a small region from a low-resolution cosmological volume, the field galaxies beyond a certain distance are unresolved. Additionally, because the zoomed-in volume covers only a fraction of the cosmological volume, it fails to sample the entire field population. Although not shown here, the median from just the detectable field galaxies in the paired suite agrees well out to $\sim$1.5\,Mpc, beyond which they too fail to capture the observed distribution. As discussed previously, we use the properties of the field galaxies as estimated from the more massive host in the paired suite to avoid overcounting.


We show the present-day cumulative stellar-mass functions (SMFs) of the streams and satellites with estimated detectability constraints in Figure \ref{fig:z0_SMFs}. In the top row, we show the SMFs constructed using only the detectable substructures and in the bottom row, we show the total SMFs obtained from the simulations. The thin lines show the SMFs from each halo, while the thick lines and the shaded regions show their median and 68\% halo-to-halo scatter. The blue and orange lines are from the isolated and paired suites, respectively. We compare the SMFs of the simulated and detectable streams in the left column. We find that most of the low-mass streams ($M_\star \lesssim 10^{7.5} \, M_\odot$) are undetectable. In the bottom row of the middle column, we show the SMFs constructed using only the substructures identified as satellites in the simulation. But in the top row, the detectable substructures also include the streams without detectable tidal tails (only the bound progenitor is detectable). We find that the simulated and detectable SMFs are similar, which means the undetectable intact satellites are largely compensated for by the streams detectable as satellites. In the right column, we show the combined SMFs of the streams and satellites. We find that the simulated and detectable SMFs of the combined population are similar to each other. While constructing the SMFs for the detectable substructures, we use their total stellar masses from the simulation by assuming perfect reconstruction of the SFH from only the stars in the detectable regions. Hence, our estimates of the present-day SMFs show the ideal scenario where we are able to extract the total present-day stellar mass of each detectable substructure. In future work, we will explore the feasibility of this assumption by modeling the CMDs to reconstruct the SFHs.


\subsection{Size-Luminosity Relation}
\label{subsec:size_lum_relation}

\begin{figure}[ht!]
    \centering
    \includegraphics[width=\linewidth]{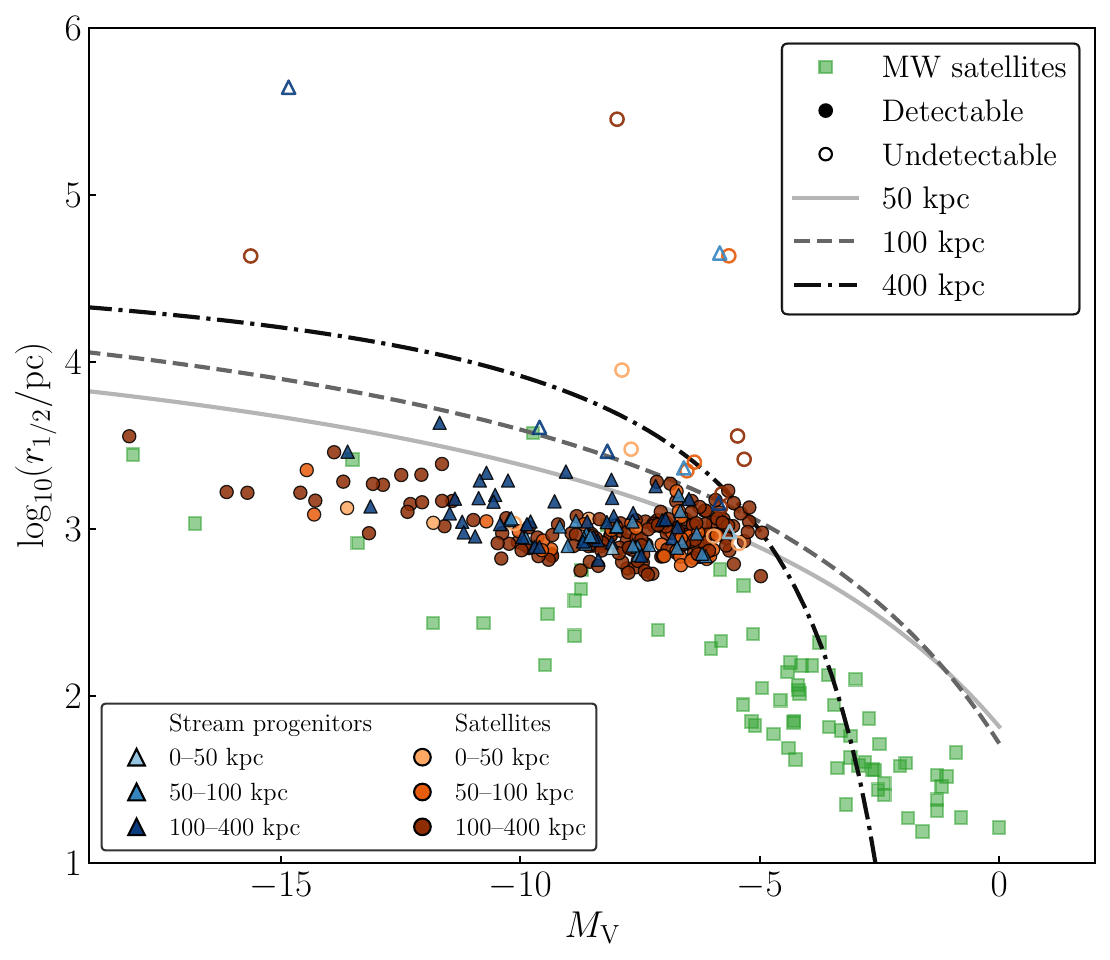}
    \caption{Satellite detectability for LSST, estimated using the analytical approximation from \citet{Tsiane_2025}. Luminosity $M_{\rm{V}}$, half-light radius $r_{1/2}$, and distance at $z = 0$ are computed from the mock observations as described in Section \ref{subsec:intact_gal_detectability}. The filled and open markers represent detectable and undetectable satellites, respectively. The blue triangles show the stream progenitors, the orange circles show the intact satellites, and the green squares show the observed MW satellites taken from LVDB. Lighter to darker shades represent the three different distance bins of 0--50, 50--100, and 100--400 kpc. The majority of our mock intact satellites and stream progenitors are detectable, with the faint, extended (low-surface-brightness) systems being undetectable and also showing a deviation from the observed trend due to resolution limits. The curves represent 50\% detectability analytical approximation at different distances.}
    \label{fig:Mv_rhalf}
\end{figure}

We compare our estimated luminosity and size for the intact galaxies and bound stream progenitors with the observed dwarf galaxies in the MW and LG. We use the observed V-band magnitude ($M_{\rm V}$) and half-light radius of the MW satellites and local field ($<3$ Mpc) galaxies from LVDB. Figure \ref{fig:Mv_rhalf} shows the $r_{\rm 1/2}$--$M_{\rm V}$ relation for our mock satellites (orange), stream progenitors (blue), and observed satellites of the MW (green). The filled markers show the detectable galaxies, while the open ones show the undetectable galaxies according to our satellite detectability scheme discussed in Section \ref{subsec:intact_gal_detectability}. For illustrative purposes, we also show the 50\% detectability curves at distances of 50\,kpc, 100\,kpc, and 400\,kpc, using the analytic relation from \citet{Tsiane_2025} for LSST.

We see good agreement between the mock and observed size-luminosity relation for brighter ($M_{\rm V} \lesssim -6$) and larger ($\log_{10}(r_{\rm 1/2}/\rm pc) \gtrsim 3$) galaxies. This suggests our ellipsoidal Plummer profile fits are a good approximation in these regions. The agreement with the observed size-luminosity indicate similar conclusion as in \citet{Klein_2024}, showing that the half-light radii from synthetic observations are better estimates for the observed sizes than particle-based half-mass radii or half-light radii assuming a constant mass-to-light ratio. Hence, even though FIRE-2 galaxies appear ``puffier'' than observed galaxies when using the particle-based size, our synthetic observations alleviate that discrepancy within the resolution limit. The tidally transformed stream progenitors tend to be slightly larger than other satellites at the same $M_V$. While there exist some fainter and smaller observed galaxies, our simulation resolutions are not high enough to probe this regime. Furthermore, our mock galaxies start to deviate from the observed trend around $M_{\rm V} \gtrsim -6$ as they approach the simulation resolution limits ($M_\star \approx 10^{5}\,M_\odot$ or $\approx20$ star particles). These galaxies are also the ones that comprise most of the undetectable fraction.

\begin{figure}[ht!]
    \centering
    \includegraphics[width=\linewidth]{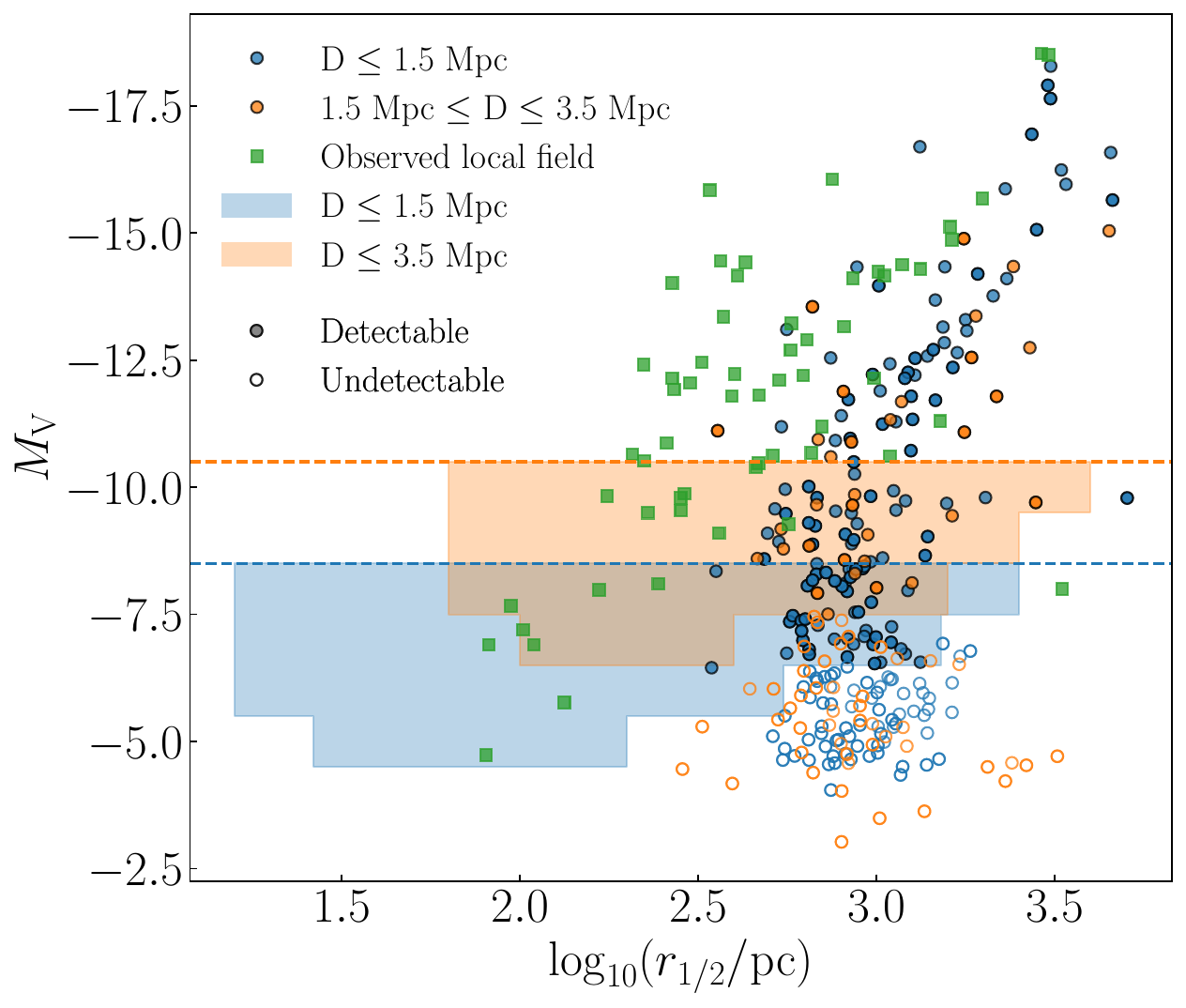}
    \caption{Field galaxy detectability for LSST, estimated using the 50\% detectability regions from \citet{Mutlu-Pakdil_2021}. The properties are calculated in the same way as for the satellites. The filled and open circles represent detectable and undetectable field galaxies respectively. The green squares represent the local field dwarfs taken from the LVDB. The blue and orange shaded regions show the 50\% detectability regions at a distance of 1.5 Mpc and 3.5 Mpc, respectively. The blue and orange markers similarly represent our mock field galaxies at a distance of $\rm{D} \leq 1.5$ Mpc and $1.5 < \rm{D} \leq 3.5$ Mpc, respectively.}
    \label{fig:Mv_rhalf_fields}
\end{figure}

In Figure \ref{fig:Mv_rhalf_fields}, we also compare the size-luminosity distribution of our mock field galaxies to observed dwarf galaxies in the local field taken from the LVDB. The local field dwarfs in the LVDB reside outside both the MW and M31, and within a distance of $\sim 3$ Mpc. Similarly, we consider the galaxies outside the virial radius of the main (most massive) host galaxy in the isolated suite as field galaxies. As discussed in Section~\ref{subsec:det_stream_gal_results}, we use the properties as estimated from the more massive host in case of the paired suite. 

\begin{figure*}[ht!]
    \centering
    \includegraphics[width=\linewidth]{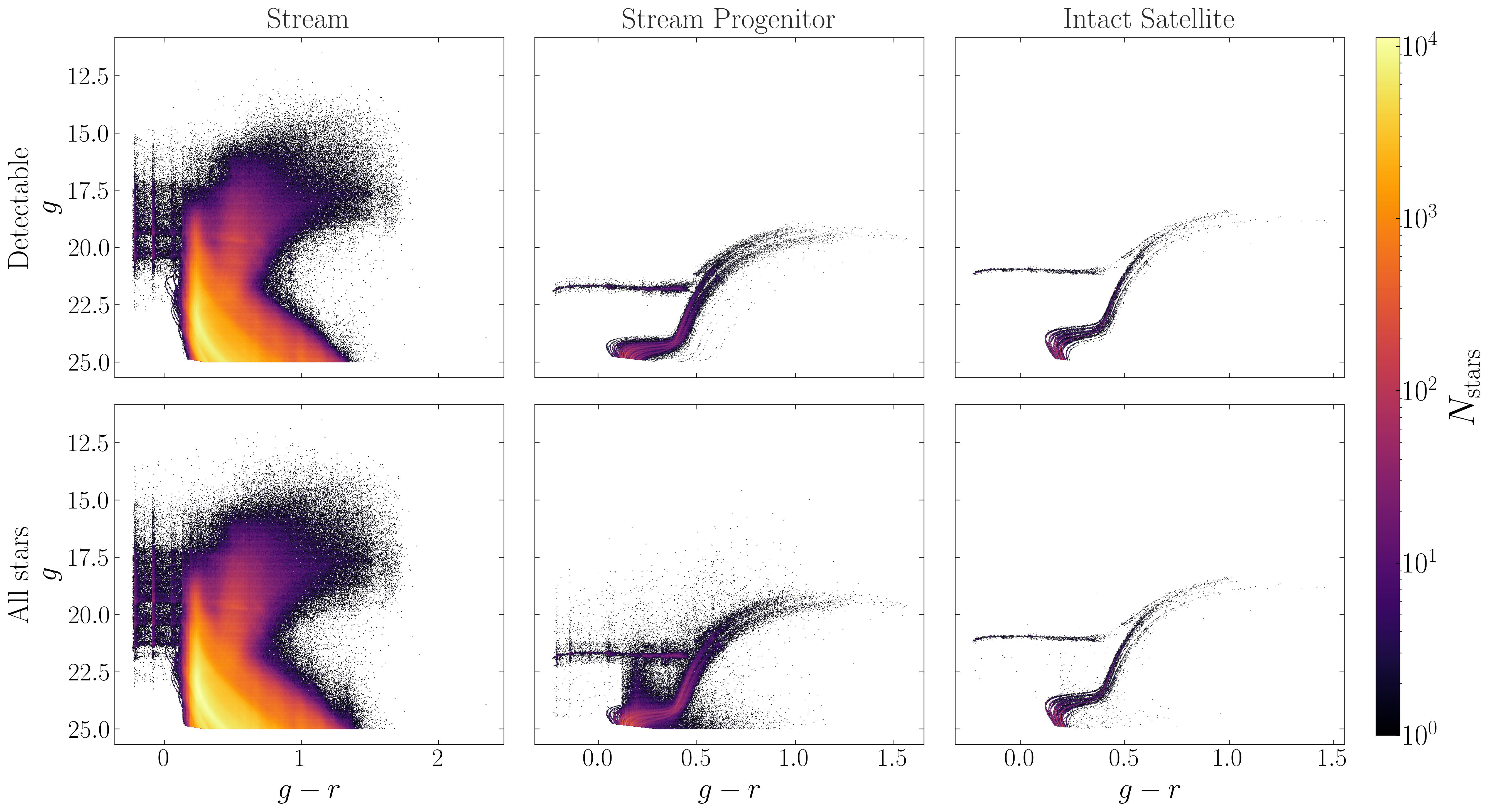}
    \caption{Examples of color-magnitude diagrams (CMDs) constructed using stars in the detectable regions (top) and all the stars within the magnitude limits of LSST (bottom).  \textit{Left}: a stream detectable with tidal tails. \textit{Middle}: a stream progenitor which is detectable as a satellite (no detectable tidal tails). \textit{Right}: a surviving satellite. In all three examples, most of the CMD can be recovered using stars from the detectable regions and hence puts confidence in SFH reconstruction from CMD modeling. Our CMDs do not include photometric errors which leads to the isochrones being apparent in this figure.}
    \label{fig:CMD}
\end{figure*}

The blue and orange points show the field galaxies within a distance of $\rm{D} \leq 1.5$ Mpc and $1.5 < \rm{D} \leq 3.5$ Mpc, respectively, from the observer. The blue and orange shaded regions show the 50\% detectability regions for distance from the observer within 1.5 Mpc and 3.5 Mpc, respectively, taken from \citet{Mutlu-Pakdil_2021}, for LSST. The filled and open circles show the detectable and undetectable galaxies, respectively according to the 50\% detectability region. As in \citet{Mutlu-Pakdil_2021}, we also classify any field galaxy above the brightest limit of the regions as detectable. Here too, we see the impact of resolution; our mock field galaxies start to deviate from observations around $(\log_{10}(r_{\rm 1/2}/\rm{pc}), \, M_{\rm V}) \approx (3, \, -6)$. However, almost all of these galaxies are below the 50\% detectability regions. Given that our systems tend to be slightly lower in surface brightness than the observed field galaxies, we can consider our analysis as a lower limit on the number of detectable field galaxies.

\subsection{Color-Magnitude Diagrams}
\label{subsec:CMD}

In order to estimate the SFHs and UV luminosities at earlier redshifts, observers use the resolved color-magnitude diagram (CMD) of a galaxy. We show example CMDs in the LSST \textit{g} and \textit{r} bands for some of our mock substructures in Figure \ref{fig:CMD}. As discussed in Section \ref{subsec:run_py-ananke}, these CMDs do not include LSST photometric uncertainties, extinction, or reddening, so the isochrones used to sample the stellar populations are still apparent in our results. Including photometric errors will blur these features. The top row shows the CMD constructed using the synthetic stars only in the detectable regions (see Section \ref{sec:detectability}), while the bottom row includes all the synthetic stars brighter than the selected LSST apparent magnitude cutoffs ($g < 25$ and $r < 24.7$). The left column is from a stream with detectable tidal tails, the middle column is from a stream progenitor with no detectable tails, and the right column is from an intact satellite. For the stream detectable with tidal tails, we include the stars within the region whose average surface brightness is $\mu_{g} \leq 34 \, \rm{mag \,\, arcsec^{-2}}$. For the stream progenitor (without detectable tidal tails) and satellites, we include all the stars within $4r_{\rm{1/2}}$ as detectable. As discussed previously, this is also the region we use to mask any bound progenitor of a stream. We find that the detectable parts of these substructures generally contain enough stars to map out the full CMD within LSST magnitude limits, but for some of the streams detectable as satellites we might lose the information from the fainter stars if they reside in the tidal tails. This is important when testing the accuracy of the reconstructed SFHs via CMD fitting. However, we show these CMDs only to illustrate the usefulness of mock catalogs and do not perform any CMD fitting to reconstruct the SFHs, as it is beyond the scope of this paper. In future data releases, we will include photometric errors and fit the CMDs using codes like \textsc{MATCH} \citep{Dolphin_2002} to get more accurate ``observed'' SFHs.

\begin{figure*}[ht!]
    \centering
    \includegraphics[width=\linewidth]{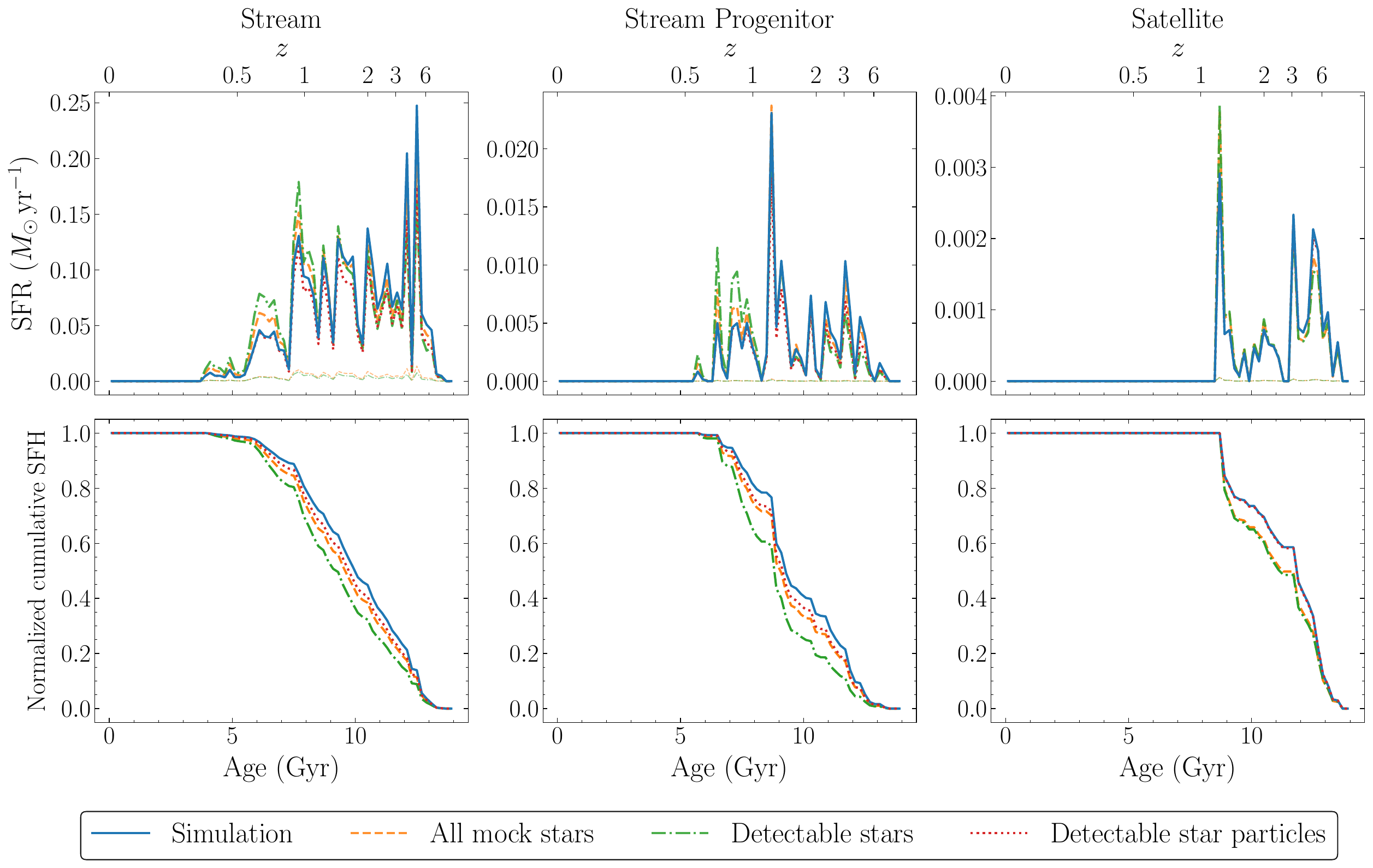}
    \caption{Examples of the star-formation histories (SFHs) for the same three substructures as in Figure \ref{fig:CMD}, constructed from the masses and ages of the stars. The blue solid curves show the simulated SFHs from the parent star particles, the orange dashed curves show the SFHs computed using all the mock stars within the apparent magnitude limits, the green dashed-dotted curves show the same computed using stars only in the detectable regions, and the red dotted curves show the simulated SFHs from parent star particles of detectable mock stars. The top row shows the star-formation rates as a function of age, and the bottom row shows the cumulative SFHs normalized to one. The mock SFHs in darker shades are renormalized using the method described in Section \ref{sec:SFHs}, while the non-renormalized versions are shown in lighter shades. The simulated SFHs are reasonably reproduced by the SFHs constructed from detectable stars. See text for more details.}
    \label{fig:SFHs}
\end{figure*}

\section{Star-Formation Histories}
\label{sec:SFHs}

SFHs from resolved photometric observations of the stellar populations in nearby galaxies are the key to inferring the UVLFs at $z \gtrsim 6$ through the near-far technique. In order to correctly reconstruct the UVLFs, it is important to accurately reconstruct the SFHs as well. However, due to stellar evolution combined with complex merger histories, it is difficult to get the complete SFH. Furthermore, detectability of the galaxy plays a crucial role in how accurately we can reconstruct the complete SFH. Hence, we compute the SFHs from our mock observations using only detectable stars and compare them against the ``true'' SFH from the simulations. We note that we do not use a CMD fitting approach like observers to reconstruct the SFHs, and hence our detectable stars encode the age and mass of their parent star particle with 100\% precision. Hence, the results shown here are for best scenarios when we have accurate and precise information of the age and mass of stars.

In Figure \ref{fig:SFHs}, we show as an example the SFHs of the same stream, stream progenitor, and intact satellite shown in Figure \ref{fig:CMD}. The blue curves show the SFHs computed directly from the entire set of simulation star particles, the orange curves include SFHs from all the mock stars within LSST magnitude limits, the green curves show the SFHs from all the stars within the detectable regions as discussed in Section \ref{subsec:CMD}, and the red curves show the SFHs from all the detectable star particles which we identify by uniquely matching the detectable mock stars to their parent star particles. In the top row, we show the differential star-formation rate (SFR) as a function of age (or redshift). We renormalize the mock ``observed'' SFRs (orange and green) to a total mass comparable to the simulated SFH (blue), since the total mass of the mock stars, whether at present day (\texttt{mact}; see Appendix \ref{appendix:mock_data}) or at formation (\texttt{smass}), is well below the actual total stellar mass of the substructure (faint orange and green curves in the top row of Figure \ref{fig:SFHs}). This is mainly because a substantial portion of the stellar mass falls below the magnitude limits imposed on our survey. Since we do not perform any CMD modeling, we cannot extrapolate the total stellar mass or luminosity of the simulated substructure from the mock stars alone. Instead, we calculate the total mass of the substructure using the formation mass of the star particles, which is what we use to calculate the SFH. Then, we calculate the total formation (or ZAMS) mass of the mock stars, provided by the \texttt{smass} quantity, in both the magnitude-limited and detection-limited cases. Mock SFRs scaled by the ratio of these two quantities are shown as the brighter orange and green curves, and are in good agreement with the ``simulation'' SFRs (blue). While this rescaling sets the total mass, it does not change the shape of the SFH, which depends on the fraction of each star particle's formation mass that ends up in detectable mock stars. Since this fraction varies strongly with the age of the star particle (see below), a single, age-independent factor cannot correct for it, and the mock SFRs overestimate later star formation and underestimate early star formation. However, this discrepancy is not prohibitively large for our calculation of the UV luminosities in most cases, as we will show in Section \ref{sec:UVLFs}. To replicate what CMD modeling using detectable stars would recover ideally, we also show the SFRs computed from only the detectable star particles (red). We find excellent agreement with the ``simulation'' SFRs from the entire system, particularly in case of the intact satellite, under the assumption that we get precise age and formation mass of the stars.

In the bottom row, which shows the cumulative SFH normalized to unity as a function of age, the relative shape of the ``true'' simulated SFH (blue) is largely recovered by the mock ``observed'' SFHs (orange and green). The remaining differences between the shape of the mock and simulated SFHs of the systems shown in this example are not primarily due to the missing stellar populations in the undetectable regions, as the SFHs derived from only the detectable star particles (red) closely match that of the entire system (blue). This shows that the detectable stars sample the full range of stellar ages. Instead, the differences arise mainly from our methodology to calculate the mock SFHs. The number of detectable mock stars per unit formation mass is much higher for younger star particles than for older ones. This is because in older populations, only evolved stars in the red-giant branch or near the main-sequence turnoff are detectable, whereas younger populations have more numerous, massive luminous stars. Summing the formation masses (\texttt{smass}) of the detectable stars in each age bin therefore underestimates earlier star formation relative to later star formation. CMD fitting of the detectable stars, which infers the total mass formed at each age while accounting for this age-dependent selection, should therefore recover the SFH of the entire system to a good extent. 

Across the full set of mock catalogs, the trends depend on the classified substructure. For streams detectable with tidal tails, the mock SFHs agree well in shape with the simulated SFHs in all but a few cases. For the streams detectable as satellites, the mock SFHs are also generally representative of the entire stream. This suggests that the bound progenitors in these systems span the entire stellar population of the stream. This could be possible if the disrupted system has gone through multiple pericentric passages leading to mixed stellar populations between the bound component and the tidal tails. Therefore, such systems would provide unbiased estimates of the high-$z$ UV magnitudes. The intact satellites show the largest discrepancies, with several cases in which the shape of the mock SFH differs substantially from the simulated one (see Figure~\ref{appendix_fig:SFH} for examples from another set of stream, stream progenitor, and satellite). However, when we reconstruct the SFHs from the detectable star particles (red), the satellites show excellent agreement with the simulation SFHs (blue). As Figure~\ref{appendix_fig:SFH} shows, the largest discrepancies occur in systems whose star formation continues to the present day, where younger populations with massive detectable stars have much larger contributions to the SFH, relative to the older populations. This shows that our SFH reconstruction method using just the \texttt{smass} quantity cannot reproduce the ``simulation'' SFHs very well in these cases. In future work, we will derive SFHs through full CMD fitting to confirm this.

\begin{figure*}[ht]
    \centering
    \includegraphics[width=\linewidth]{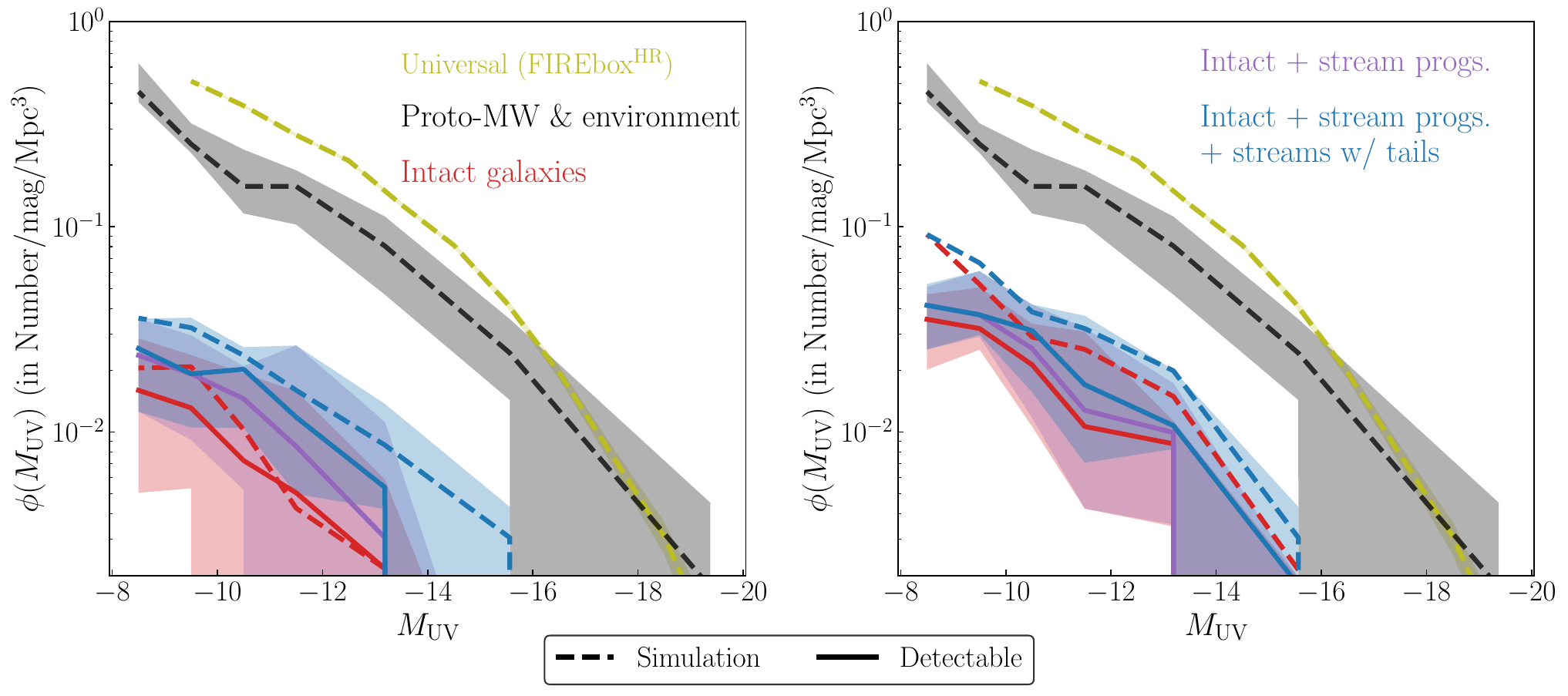}
    \caption{Reconstructed ultraviolet luminosity functions (UVLFs) at $z = 7$. The simulated UVLFs are represented by the dashed curves, while the solid curves are reconstructed by using only detectable substructures. All the curves show the median across the 13 hosts, while the shaded regions denote the 68\% halo-to-halo scatter. The red curves include the progenitors of only the intact galaxies --- satellites (left) and satellites + fields (right). The purple curves include the intact galaxies and streams detectable as satellites, while the blue curves also include the streams. The black curves represent the UVLF of the region containing the proto-MW and its environment. The olive curve shows the UVLF from FIREbox$^{\rm{HR}}$, which we use as the benchmark for the universal UVLF. Including the streams leads to a better reconstruction of the proto-MW UVLFs, both in terms of the normalization and shape. See text for details.}
    \label{fig:UVLFs}
\end{figure*}

\section{Ultraviolet Luminosity Functions}
\label{sec:UVLFs}

The main goal of the near-far technique is to probe the faint end of the UVLF during reionization. In \citet{Kundu_2026}, we explored how accurately we can recover the UVLFs of the proto-MW/LG at $z \gtrsim 6$ by adding the progenitors of present-day disrupted galaxies along with the intact galaxies in the \emph{all-progenitors} and \emph{fossil-record} scenarios. The \emph{all-progenitors} scenario uses the full merger history information to account for the fact that a present-day substructure can have multiple high-$z$ progenitors, while the \emph{fossil-record} scenario combines the total stellar mass from all the progenitors of each present-day substructure into one. In this work, we only focus on the \emph{fossil-record} scenario, since this is the main assumption of the near-far technique and can also recover faint-end slopes depending on the selection criteria of present-day galaxies (see \citealt{Kundu_2026} for details). Furthermore, we only add the stellar streams to our present-day sample of intact galaxies and ignore the phase-mixed galaxies, as we are not estimating their detectability in this work.

To calculate the rest-frame UV magnitude ($M_{\mathrm{UV}}$) at $1500\,\rm{\AA}$ for the progenitor of each simulated present-day substructure, we track its constituent star particles to $z \gtrsim 6$ and select the ones that already formed by that redshift. Then, using their masses and ages at that redshift, along with their metallicities, we compute the total $M_{\mathrm{UV}}$ of the progenitor, following the approach in \citet{Sun_2023}. In case of the mock observations, we follow a slightly different approach. We select all the ``mock'' or ``detectable'' stars that formed by our redshift of interest, take their formation mass (\texttt{smass}) and multiply by the factor that we calculated in Section \ref{sec:SFHs}. The rest of the approach is the same as the one we used in case of the ``simulated'' star particles. Once we have $M_{\mathrm{UV}}$ for the progenitor of each present-day substructure, we compute the differential UVLFs following the convex-hull volume calculation method described in \citet{Kundu_2026}.

\begin{figure*}[ht]
    \centering
    \includegraphics[width=\linewidth]{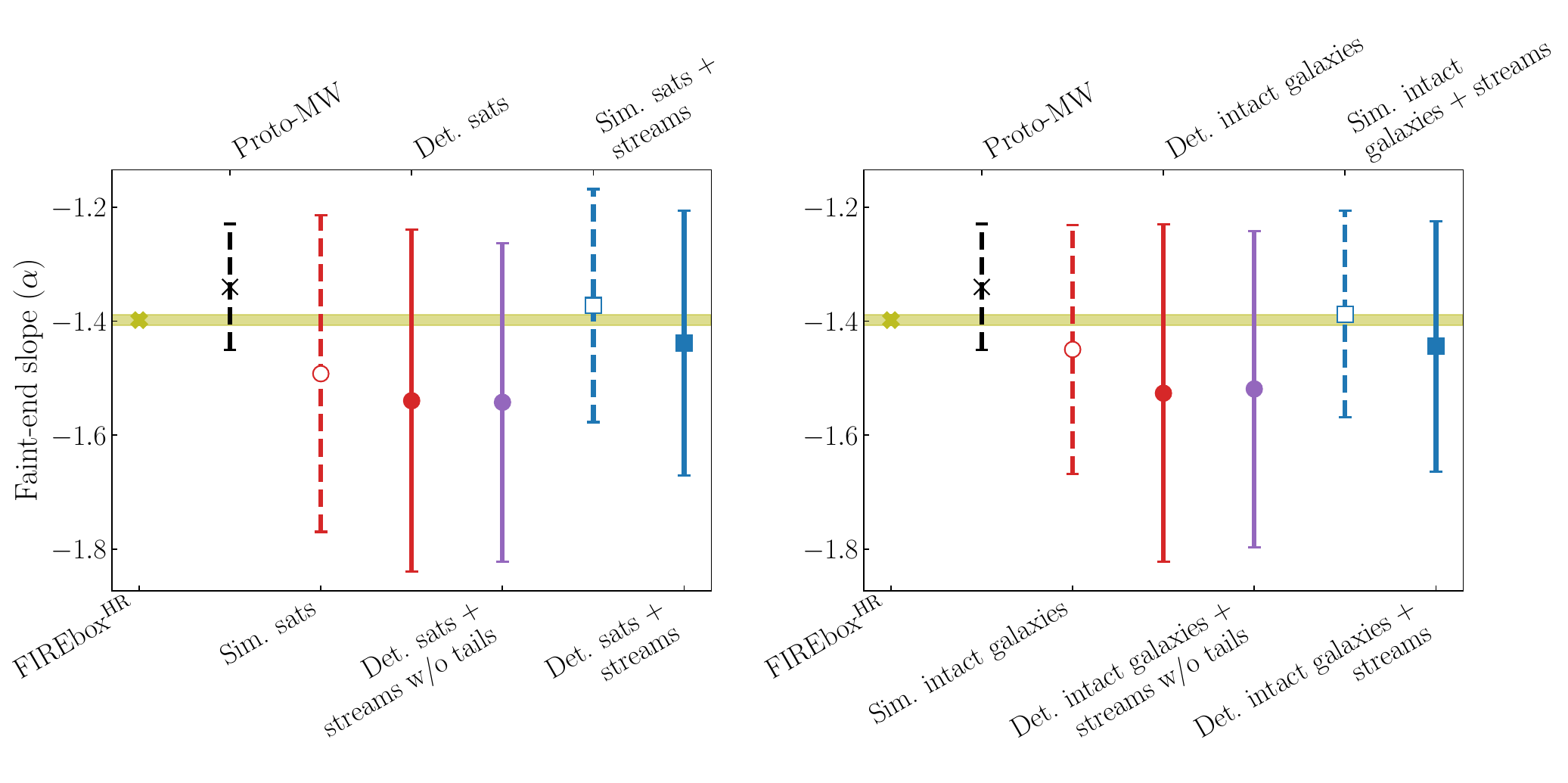}
    \caption{Faint-end slopes of the UVLFs at $z = 7$ estimated using the progenitors of the present-day simulated substructures (dashed lines) and their detectable stars (solid lines). The markers indicate the central estimate across the 13 hosts, while the error bars indicate the halo-to-halo scatter. The proto-MW slope is denoted in black and the slopes using the progenitors of the intact galaxies (only satellites on left and satellites + fields on right) are shown in red. The slope estimated using detectable satellites along with streams with undetectable tidal tails (detectable as satellites) is shown in purple. The best-fit universal faint-end slope from FIREbox$^{\rm{HR}}$ is denoted by the olive point and the shaded region shows the 68\% CI (not a halo-to-halo scatter). Adding the streams (blue) improves the accuracy in estimating the proto-MW and the universal faint-end slope and also reduces the halo-to-halo scatter. See text for details.}
    \label{fig:UVLF_slopes}
\end{figure*}

In Figure \ref{fig:UVLFs}, we show the reconstructed differential UVLFs at $z=7$ from the fossil record using simulated star particles and present-day detectable mock stars. In the left panel, we consider only satellites as intact galaxies, while in the right panel we also add the field galaxies to the list of intact galaxies. We separate the two in order to check how well we can reconstruct the high-$z$ UVLFs using only the MW satellites. Furthermore, in the paired suite, the field galaxies are common to both the hosts. This can lead to a bias in the halo-to-halo scatter as these galaxies are drawn from the same distribution. The lines show the median across the 13 hosts, while the shaded regions (only shown for the detectable cases) show the 68\% halo-to-halo scatter. The dashed blue and red curves show the reconstructed UVLFs from the progenitors of present-day simulated intact galaxies (satellites on the left panel; satellites and field galaxies on the right panel) with and without the streams, respectively. The solid red curve shows the reconstructed UVLF using only the detectable intact galaxies. The solid purple curve includes the streams detectable as satellites, while the solid blue curve also adds detectable streams. We compare all the reconstructed UVLFs with that of the proto-MW/LG as shown by the black dashed curve and the UVLF from the FIREbox$^{\rm{HR}}$ (olive) large-volume simulation \citep{Feldmann_2025}, which we use as the benchmark for the universal UVLF during EoR. The olive shaded region shows the 68\% confidence interval (CI) obtained via bootstrapping, and not a halo-to-halo scatter. The difference between the proto-MW/LG and FIREbox$^{\rm{HR}}$ UVLF is due to the methodological difference when calculating the distributions. We refer the reader to \citet{Kundu_2026} for detailed discussion. Adding the streams detectable as satellites improves the normalization of the UVLF of detectable intact galaxies by a factor of $\sim$1.5. This shows that the ``streams as satellites'' population can contribute significantly to the reconstructed UVLF at high redshifts. Furthermore, when comparing simulations with observations, these objects can be mistakenly compared with the satellite population of the simulations.  The near-far reconstruction will need to account for the missing stellar populations in the tidal tails for cases where the bound components are biased tracers of the entire population. However, in Section~\ref{sec:SFHs}, we find the mock SFH for most of these systems agrees well with the simulated SFH for the entire stream. It is mostly the actual satellite galaxies that have larger differences between the mock and simulated SFHs. We emphasize that this could be due to our methodology for computing the SFHs and needs further investigation. Adding the streams with detectable tidal tails, we improve upon the normalization at $M_{\mathrm{UV}} \lesssim -10$, which is consistent with what we found in \citet{Kundu_2026}. The more massive streams are easily detectable with tidal tails and hence amplify the brighter end of the reconstructed UVLF at high redshifts. The difference in the normalization of the black and blue curves is due to the progenitors of the present-day phase-mixed galaxies and the main host galaxy, which we have not considered here. The shape of all the curves (both simulated and detectable) is qualitatively similar to that of the proto-MW/LG and the universal UVLF down to $M_{\mathrm{UV}} \approx -10$. Furthermore, adding the stream population improves the normalization by at least a factor of $\sim2$. Since LSST will detect more of these systems, they should be included while reconstructing the reionization-era UVLFs using the near-far technique.

For a quantitative comparison, we calculate the slope at the faint end of the UVLFs assuming a Schechter function \citep{Schechter_1976} and performing a Bayesian modeling using the \texttt{PyMC} package \citep{PyMC}, as described in \citet{Kundu_2026}. Figure \ref{fig:UVLF_slopes} shows the faint-end slopes at $z = 7$ for both the simulated and detected cases. The points show the central estimate of the slope across all the 13 halos and the error bars indicate the bootstrapped halo-to-halo scatter (see \citealt{Kundu_2026} for details). The olive point and the shaded region show the best-fit value and uncertainty (not halo-to-halo scatter) in the faint-end slope from the FIREbox$^{\rm{HR}}$ UVLF. The black points show the faint-end slope of the proto-MW UVLF, while the red points show the slope from only the intact galaxy progenitors (left panel includes only satellites; right panel also includes the field galaxies). The blue points show the slope when streams are also added to the sample of present-day intact galaxies. The dashed error bars with open markers indicate the slopes estimated from the progenitors of the simulated substructures, while the solid error bars with filled markers denote the slopes calculated from the progenitors of only detectable substructures. The purple point includes detectable satellites and streams without tidal tails (streams detectable as satellites). While this is closer to the slopes measured observationally in the near-far technique, we can see that the accuracy is very similar to using only satellites. However, we see a decrease in the scatter which leads to a better constraint on the proto-MW slope. Adding the detectable streams yields more accurate estimates of both the proto-MW and universal faint-end slopes. Additionally, we reduce the halo-to-halo scatter by $\approx33\%$. As shown in the right panel, adding the detectable field galaxies to the sample of present-day intact galaxies provides similar slopes.
However, our estimates are also biased by the common distribution of field galaxies in the paired hosts. Adding detectable streams from LSST can thus substantially improve both the accuracy of the recovered faint-end slope of the proto-MW UVLF during EoR and its robustness against halo-to-halo scatter.

\section{Summary}
\label{sec:summary}

In this work, we generate and publicly release 13 mock LSST catalogs of resolved stars in both intact and disrupted galaxies around MW-mass hosts in the FIRE-2 cosmological zoom-in simulations (7 isolated MW analogs and 6 hosts from 3 LG-like pairs) using \textsf{py-ananke}. The catalogs provide astrometry, photometry in the six LSST bands, stellar parameters, and chemical abundances for each synthetic star, along with indices linking them to their parent star particles (Appendix~\ref{appendix:mock_data}). Using these catalogs, we estimate the detectability of satellites, stellar streams, and field galaxies, identified in \citet{Kundu_2026}, and test whether including detectable stellar streams in addition to the intact galaxies improves near-far reconstructions at the faint end of the UVLF during the EoR. Our results are summarized below.


\begin{enumerate}

    \item Estimating the detectability of the tidal tails of the streams using a surface brightness limit of $\mu_{\rm{g}} \leq 34$\,mag\,arcsec$^{-2}$ and 1-year LSST magnitude limits, we find an average of $\approx$29\% of our simulated streams with $M_{\star} \gtrsim 5 \times 10^5 \, M_{\odot}$ are detectable with tidal tails (a median of three per host) across the 13 halos. Another $\approx38\%$ of the streams are detectable only through their bound progenitors and would therefore be classified as satellites. The remaining $\approx33\%$ are undetectable, and most streams with $M_\star \lesssim 10^{7.5}\,M_\odot$ fall in this category. 

    \item We estimate the detectability of the intact galaxies (satellites and fields) and bound remnants of stellar streams (stream progenitors) using analytic detectability relations in the size--luminosity ($r_{1/2}-M_{\rm{V}}$) plane. We find that an average of
    $\approx$96\% of satellites, $\approx$95\% of bound stream progenitors, and $\approx$64\% of field galaxies (out to $\sim$3.5\,Mpc) with $M_\star \gtrsim 10^{5}\,M_\odot$ are detectable, with the numbers decreasing to $\approx$59\%, $\approx$89\%, and $\approx$42\%, respectively, for $M_\star \gtrsim 10^{4.5}\,M_\odot$ owing to resolution limits. Our mock galaxies follow the observed size--luminosity relation of MW satellites and local field dwarfs for $M_V \lesssim -6$.

    \item The cumulative radial distribution of detectable satellites alone underpredicts the observed MW satellite distribution, which lies within the 68\% halo-to-halo scatter only beyond $\sim$220\,kpc. Including streams detectable as satellites improves the radial distribution and agrees well with observations beyond $\sim$150\,kpc. When we limit our sample to substructures with $M_\star \gtrsim 10^5\,M_\odot$, the combined population agrees well with the observed distribution over the entire distance range within the halo-to-halo scatter. Our results suggest that some known MW satellites may be the bound remnants of disrupted systems. The remaining deficit within $\sim$150\,kpc might be driven by the absence of an LMC analog in most of our MW-mass hosts. Detectable field galaxies reproduce the observed local field distribution out to $\sim$800\,kpc, beyond which the zoom-in volumes become incomplete.

    \item SFHs computed from only the detectable mock stars broadly recover the simulated SFHs for most streams and streams detectable as satellites, indicating that their bound remnants typically sample the full stellar population. However, we emphasize that our reconstructions are based on 100\% precise age and mass of stars, which is not possible in the observer-approach of CMD fitting. Thus, our results are optimistic reconstructions of the SFHs. Even under this assumption, discrepancies exist, the largest of which are typically found in satellites with star formation continuing to the present day. These discrepancies are mainly due to our simplified SFH reconstruction method using the formation mass and age of the mock stars, readily available in the catalogs. They are not from missing stellar populations, as SFHs from only the detectable star particles closely match the full simulated SFHs in most cases. Hence, reconstructions based on fitting the CMDs of LSST-detectable stars can recover the SFHs to a good extent, which we will test in future work. 

    \item The qualitative shapes of reconstructed UVLFs at $z = 7$ using only detectable substructures are in good agreement with those from the simulated populations and with the proto-MW UVLF. We also find that adding the detectable streams improves the normalization of the reconstructed UVLFs --- adding the streams detectable as satellites improves the normalization by a factor of $\sim$1.5, and adding all detectable streams (including streams with detectable tidal tails) by a factor of $\gtrsim2$ around $M_{\rm UV} \sim -10$. 

    \item From our estimation of the faint-end slopes of the UVLFs at $z = 7$, we find that adding the streams with detectable tidal tails along with the intact galaxies and stream progenitors reconstructs a more accurate and precise faint-end slope of the proto-MW UVLF. Using only the intact galaxies we recover a faint-end slope steeper by $\approx$0.2 than the median proto-MW slope across the 13 halos. Adding the detectable streams reduces this offset by $\approx0.1$ and reduces the halo-to-halo scatter by $\approx$33$\%$.
\end{enumerate}

Our results suggest that LSST-detectable substructures can feasibly be used in the near-far technique to infer the high-$z$ UVLF during the EoR, and stellar streams can improve the constraints on the reconstructions, even with 1-year magnitude limits. This opens up the possibility that proper motions for stream stars derived from LSST observations \citep{Ivezic_2019,Gomes_2025} can be used to improve stream membership selections for the near-far technique, which can lead to a substantially cleaner selection. Because a substantial fraction of streams will be observed only as satellites, near-far analyses should allow for the possibility that some observed satellites are the bound remnants of larger disrupted systems, although in our simulations these remnants are generally representative of the full stellar population.

Our analysis includes some caveats. We assume perfect photometry without extinction and neglect contamination from unresolved background galaxies, which would require modeling star-galaxy separation; we expect this will have the effect of making the surface brightness cutoff slightly brighter. Our results might change depending on the actual surface brightness limit LSST will achieve. In future work, we will add photometric uncertainties to the catalogs and reconstruct SFHs through an observer-oriented approach of fitting the CMDs. We have also not estimated the detectability of the phase-mixed objects as it is beyond the scope of this work, but they are available in our mock catalogs. Adding the phase-mixed galaxies would further improve our reconstruction of the faint-end slope as we showed in \citet{Kundu_2026}. Beyond the near-far reconstructions of the high-$z$ UVLFs, these publicly available catalogs can support studies of tidal disruption, halo tracer populations such as RR Lyrae and M giants (Alvarado et al., in prep.), and the chemical abundances of intact and disrupted galaxies.

\begin{acknowledgments}
We would like to thank Robert Feldmann for providing us the FIREbox$^{\rm{HR}}$ UVLF. We would also like to thank Peter Ferguson for providing us the LSST resources and helpful discussions. We would also like to thank Yasmine Meziani and Pratik Gandhi for insightful comments which improved the quality of the manuscript. We acknowledge use of the \textsf{marmalade} cluster at the University of Pennsylvania, on which the analysis of this work was done. AK and RES acknowledge support from NSF grants AST-2007232 and AST-2307787. RES further acknowledges support from NASA awards 22-ROMAN22-0055, 22-ROMAN22-0011, and 22-ROMAN22-0013, Simons Foundation grant 1018462, and Sloan Foundation award FG-2023-20669.
\end{acknowledgments}

\software{astropy \citep{Astropy_2013, Astropy_2018, Astropy_2022},
matplotlib \citep{Matplotlib_2007},
IPython \citep{IPython_2007},
NumPy \citep{Numpy_2011, Numpy_2020}, 
SciPy \citep{Scipy_2020},
GizmoAnalysis \citep{GizmoAnalysis_2020},
HaloAnalysis \citep{HaloAnalysis_2020, Wetzel_2016},
PyMC \citep{PyMC},
healpy \citep{Healpy_2019},
mhealpy \citep{mhealpy_2022},
scikit-learn \citep{Scikit-learn_2011},
py-ananke \citep{Thob_2024},
NASA's Astrophysics Data System (ADS),
arXiv
}   


\appendix
\restartappendixnumbering

\section{Radial distribution of stream progenitors and intact galaxies with higher stellar mass cutoff}
\label{appendix_sec:radial_profile}

\begin{figure*}[h]
    \centering
    \includegraphics[width=0.5\textwidth]{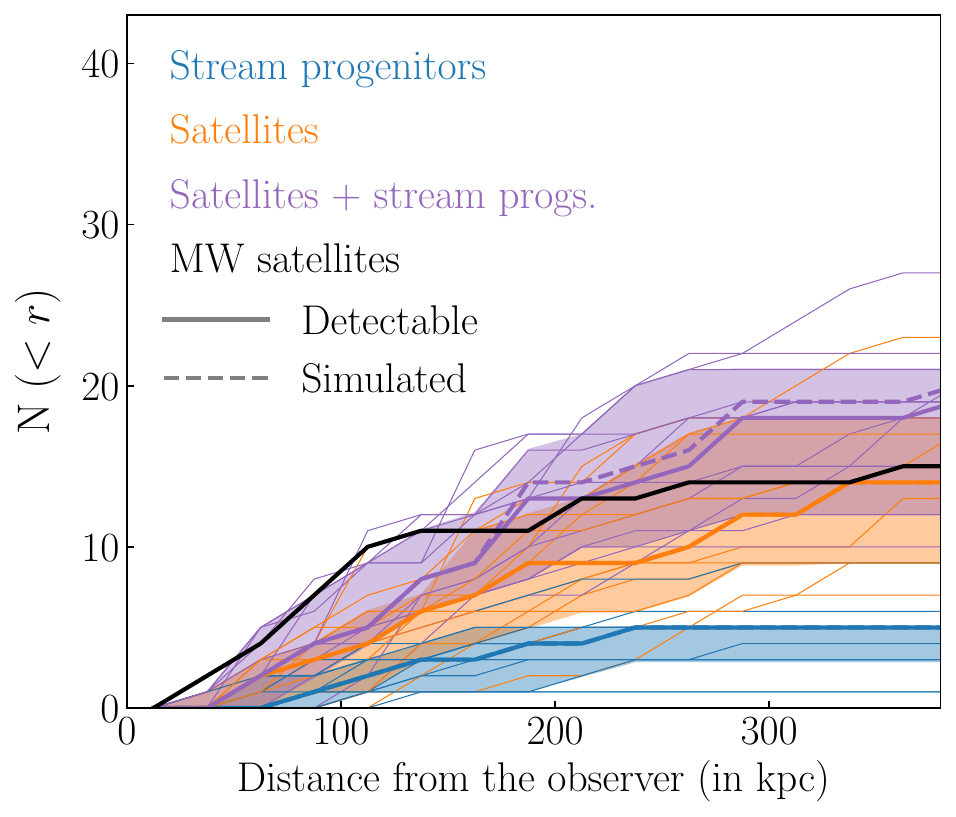}
    \caption{Radial distribution of satellites, stream progenitors, and their combined population, compared against the observed MW satellite distribution. The stellar mass cutoff is $M_\star \geq 10^5\,M_\odot$ for each population. The thick lines show the median distributions, while the thin lines show the per halo distribution. The shaded region shows the halo-to-halo scatter. The key takeaways are similar to those from Figure~\ref{fig:radial_profile}.} 
    \label{appendix_fig:radial_profile}
\end{figure*}

Figure~\ref{appendix_fig:radial_profile} shows the radial distribution of the stream progenitors and the intact satellites with $M_\star \geq 10^5\,M_\odot$.  Similar to Figure~\ref{fig:radial_profile}, we show the median distribution of the detectable populations across all the 13 host galaxies (solid lines). The shaded regions show the 68\% halo-to-halo scatter. The thin lines show the detectable populations from each of the host galaxies. The dashed curves show the radial distribution of the simulated substructures, which includes the undetectable populations. The black curve shows the distribution of observed satellites in the MW with $M_\star \geq 10^5\, M_\odot$, taken from LVDB. Similar to Figure~\ref{fig:radial_profile}, the observed MW satellite radial distribution agrees better with the distribution of combined population of detectable satellites and streams detectable as satellites, and is within the halo-to-halo scatter over the entire distance range. Additionally, we find that the distribution from the detectable systems is in excellent agreement with the simulated one for this stellar mass cutoff.

\section{Examples of large differences between mock and simulated star-formation histories}

In this section, we show a set of example SFHs where the discrepancy between the mock and simulated SFHs is large. Similar to Figure~\ref{fig:SFHs} we show the SFHs for a stream, a stream progenitor, and a satellite. The top row shows the SFRs as a function of age (or redshift) and the bottom row shows the SFHs normalized to unity. The ``observed'' SFHs from all the mock stars (orange) as well as the detectable mock stars (green) do not agree well with the ``simulation'' SFHs (blue). This is due to our SFH reconstruction method as discussed in Section~\ref{sec:SFHs}. The detectable mock stars span the entire age distribution of the simulated star particles which leads the SFH from detectable star particles to have excellent agreement with the simulated one. This suggests that CMD fitting from detectable LSST stars can recover the SFHs to a good extent.

\begin{figure*}[ht!]
    \centering
    \includegraphics[width=\textwidth]{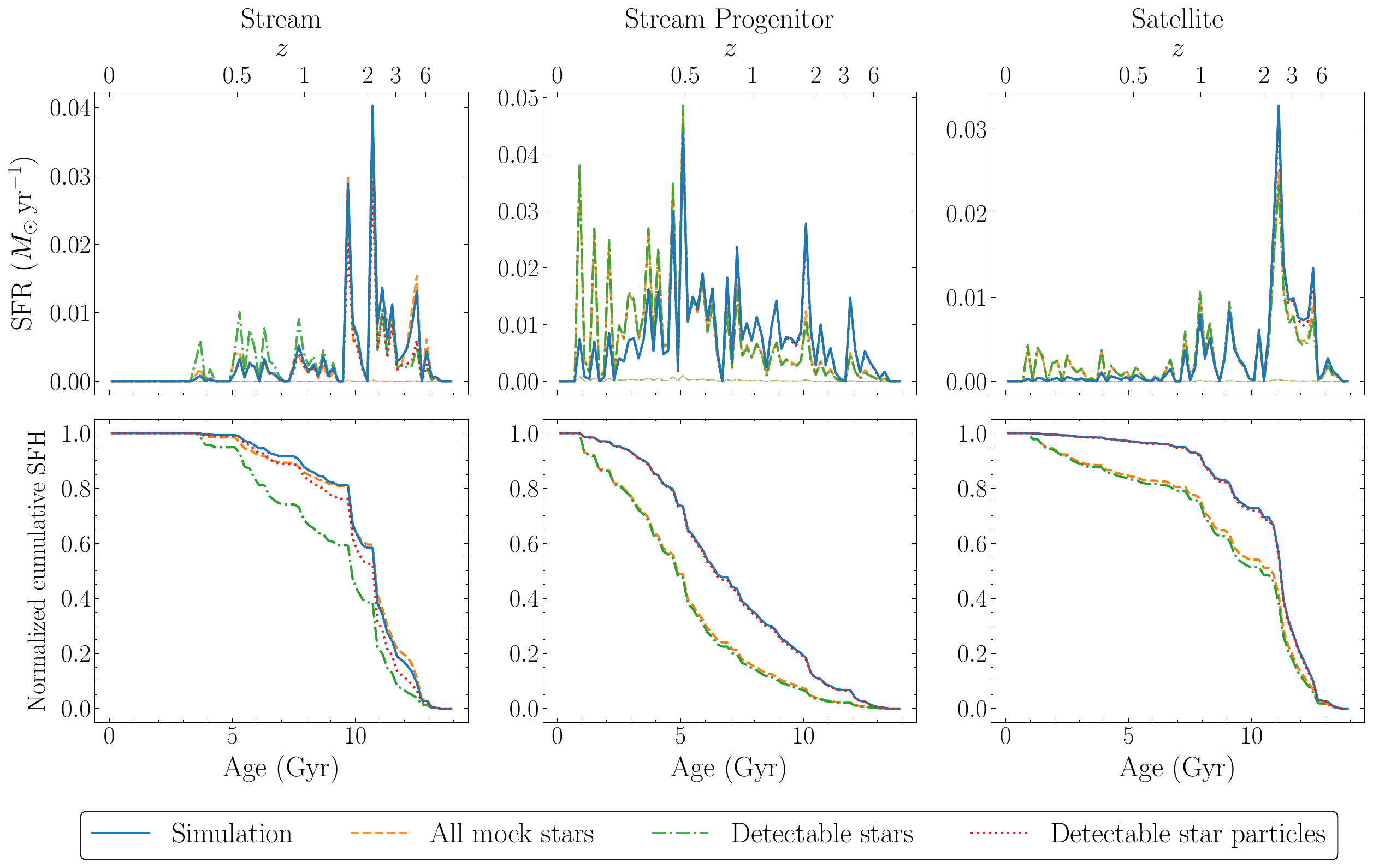}
    \caption{Examples of the star-formation histories (SFHs) showing large differences in the shape of the observed and simulated cases. The notation is the same as in Figure~\ref{fig:SFHs}. Most of these differences arise when star formation continues late to the present day. See text for details.} 
    \label{appendix_fig:SFH}
\end{figure*}

\section{Data in the mock catalogs} \label{appendix:mock_data}

In this section, we describe in detail each of the fields in the mock catalogs. For each of the 13 host galaxies, we generate synthetic stars for every intact and disrupted galaxy. A given mock observation contains multiple files partitioned according to the Healpix tessellation scheme described in Section \ref{subsec:run_py-ananke}. For a complete mock observation of a particular galaxy, the user needs to read all the partitioned files. A summary of the data products including their units is available in Table \ref{tab:mock_data}. 

\subsection{Indices} \label{subsec:indices}

Our catalog contains a number of different integers for identifying the synthetic stars and their associated generating star particle in the simulations.

\begin{itemize}

    \item \texttt{parentid} - array index of the star particle in the simulation from which a given synthetic star was generated. Synthetic stars with the same \texttt{parentid} originated from the same parent star particle in the simulation. This index can be used to get the full properties of the star particle in the simulation snapshot at $z = 0$. We also note that the star particles in FIRE have a particle ID which is not unique across the particles and we do not use them in our pipeline. Hence, \texttt{parentid} just refers to the array index of the star particle in the $z = 0$ simulation snapshot.

    \item \texttt{partid} - indicates whether a synthetic star is located at the exact phase-space coordinates of its generating star particle. The first synthetic star to be generated is always assigned the coordinates of its parent star particle and has \texttt{partid} = 0, while the rest are assigned \texttt{partid} = 1.

    \item \texttt{paritionid} - indicates the partition index of the Healpix tessellation scheme. Synthetic stars belonging to the same Healpix partition have the same \texttt{paritionid}.
\end{itemize}

\subsection{Phase-space Coordinates} \label{subsec:phase-space}

We provide different types of phase-space coordinates associated with each synthetic star. We emphasize that we do not apply observational errors to these values and hence they are the true coordinates. Future data releases may contain the error convolved versions of them.

\begin{itemize}

    \item \texttt{ra, dec} - the astrometric Right Ascension (RA) and Declination (DEC) of the synthetic star on the celestial sphere. 

    \item \texttt{px, py, pz} - the Cartesian positions of the synthetic star relative to the observer (or LSR in this work) position. 

    \item \texttt{vx, vy, vz} - the Cartesian velocities of the synthetic star relative to the observer (or LSR in this work) velocity.

    \item \texttt{rad} - the radial distance of the synthetic star to the observer.

    \item \texttt{vr} - the radial velocity of the synthetic star to the observer.

    \item \texttt{dmod} - the distance modulus of the synthetic star.

    \item \texttt{dform} - the formation distance of the synthetic star. FIRE simulations track the distance of the star particles from the center of the host galaxy at their time of formation. We provide this as an input to the pipeline and hence synthetic stars originating from the same parent star particle will have the same formation distance.

    \item \texttt{pi} - the parallax of the synthetic star. It is related to the distance modulus via 

    \begin{equation}
        \frac{\texttt{pi}}{\rm{mas}} = \frac{1000}{10^{\frac{\texttt{dmod}}{5} + 1}}
    \label{eq:parallax}
    \end{equation}

    \item \texttt{glat, glon} - the Galactic latitude and longitude of the synthetic star on the celestial sphere.

    \item \texttt{mura, mudec} - the astrometric proper motions in the direction of RA and DEC, respectively. 

    \item \texttt{mul, mub} - the astrometric proper motions in the direction of Galactic longitude and latitude respectively. 
\end{itemize}

\subsection{Photometry} \label{subsec:photometry}

We have two types of photometric magnitudes for the LSST bands in our mock catalogs:

\begin{itemize}
    \item \texttt{lsst\_<filtername>\_Intrinsic} - absolute AB magnitude of the synthetic star, where \texttt{<filtername>} is any one of the LSST \textit{ugrizy} bands.

    \item \texttt{lsst\_<filtername>} - apparent AB magnitude of the synthetic star. For now, this does not include galactic extinction or photometric errors, which future data releases will contain. Hence, the apparent and absolute magnitudes differ only by the distance modulus.
\end{itemize}

\subsection{Stellar Parameters} \label{subsec:stellar_params}

We provide the values of some stellar parameters for the synthetic stars in our catalog, obtained from the isochrone models described in Section \ref{subsec:run_py-ananke}. They do not have any observational errors associated with them.

\begin{itemize}
    \item \texttt{mact} - present-day stellar mass of the synthetic star after stellar evolution.

    \item \texttt{smass} - mass of the synthetic star when it was on the zero-age main sequence (ZAMS).

    \item \texttt{mtip} - mass of the synthetic star at the tip of the giant branch for its given age and metallicity. All stars originating from the same parent star particle have the same \texttt{mtip} value.

    \item  \texttt{teff} - effective temperature of the synthetic star in $\rm K$.

    \item \texttt{lum} - stellar luminosity of the synthetic star in $\rm L_{\odot}$.

    \item \texttt{grav} - $\log_{10}$ surface gravity of the synthetic star in $\rm{cm \, s^{-2}}$.

    \item \texttt{age} - $\log_{10}\rm{(age/yr)}$, obtained directly from the age of the parent star particle. All stars originating from the same parent star particle have the same age.
\end{itemize}

\subsection{Chemical Abundances} \label{subsec:chemical_abundances}

FIRE simulations track the abundances of 11 elements (H, He, C, N, O, Ne, Mg, Si, S, Ca, Fe) through the stellar yield tables and subgrid turbulent metal diffusion models (see \citet{Hopkins_2018a} for details). We forward these abundances directly to each synthetic star originating from its parent star particle. Thus, all synthetic stars spawning from the same parent star particle share the same abundances. All abundances are given relative to hydrogen and normalized to solar values \citep{Asplund_2009}. We also provide the \texttt{alpha} abundances, defined as the mean abundance using O, Mg, Si, and Ca, relative to solar. The list of chemical abundances is reported in Table \ref{tab:mock_data}.

\begin{deluxetable}{lll}
\tablecaption{Summary of the data products in the mock catalogs. \label{tab:mock_data}}
\tablehead{
\colhead{Data} & \colhead{Description} & \colhead{Units}
}
\startdata
\cutinhead{Indices}
\texttt{paritionid} & Index of the Healpix partition & \nodata \\
\texttt{parentid} & Array index of the parent star particle in the $z=0$ simulation snapshot & \nodata \\
\texttt{partid} & Flag for a star at the exact phase-space coordinates of its parent particle ($0$ or $1$) & \nodata \\
\cutinhead{Phase-space Coordinates}
\texttt{ra, dec} & Astrometric Right Ascension and Declination on the celestial sphere & deg \\
\texttt{px, py, pz} & Cartesian positions relative to the observer (LSR) position & kpc \\
\texttt{vx, vy, vz} & Cartesian velocities relative to the observer (LSR) velocity & $\rm km\,s^{-1}$ \\
\texttt{rad} & Radial distance to the observer & kpc\\
\texttt{vr} & Radial velocity to the observer  & $\rm km\,s^{-1}$\\
\texttt{dmod} & Distance modulus & \nodata \\
\texttt{dform} & Formation distance from the host galaxy center & kpc \\
\texttt{pi} & Parallax (see Equation~\ref{eq:parallax}) & mas \\
\texttt{glat, glon} & Galactic latitude and longitude on the celestial sphere & deg \\
\texttt{mura, mudec} & Astrometric proper motions along RA and DEC & $\rm mas\,yr^{-1}$ \\
\texttt{mul, mub} & Astrometric proper motions along Galactic longitude and latitude & $\rm mas\,yr^{-1}$ \\
\cutinhead{Photometry}
\texttt{lsst\_<filtername>\_Intrinsic} & Absolute magnitude in LSST band \texttt{<filtername>} ($ugrizy$) in AB system & mag \\
\texttt{lsst\_<filtername>} & Apparent magnitude in LSST band \texttt{<filtername>} ($ugrizy$) in AB system & mag \\
\cutinhead{Stellar Parameters}
\texttt{mact} & Present-day stellar mass after stellar evolution & $\rm M_{\odot}$ \\
\texttt{smass} & Zero-age main sequence (ZAMS) mass & $\rm M_{\odot}$ \\
\texttt{mtip} & Mass at the tip of the giant branch for the star's age and metallicity & $\rm M_{\odot}$ \\
\texttt{teff} & Effective temperature & K \\
\texttt{lum} & Stellar luminosity & $\rm L_{\odot}$ \\
\texttt{grav} & $\log_{10}$ surface gravity & $\rm cm\,s^{-2}$ \\
\texttt{age} & $\log_{10}$(age/yr) from the parent star particle & \nodata \\
\cutinhead{Chemical Abundances (derived from parent star particle)}
\texttt{<element>} & [X/H] abundance of element X (H, He, C, N, O, Ne, Mg, Si, S, Ca, Fe) relative to solar & dex \\
\texttt{alpha} & Mean of [O/H], [Mg/H], [Si/H], and [Ca/H], relative to solar & dex \\
\enddata
\tablecomments{Abundances are relative to hydrogen and normalized to the solar values of \citet{Asplund_2009}. Elements He, C, N, O, Ne, Mg, Si, S, Ca, \& Fe are respetively identified in the data files as \texttt{helium}, \texttt{carbon}, \texttt{nitrogen}, \texttt{oxygen}, \texttt{neon}, \texttt{magnesium}, \texttt{silicon}, \texttt{sulphur}, \texttt{calcium}, \& \texttt{feh}. Extinction, reddening, and photometric errors are not included in the apparent magnitudes.}
\end{deluxetable}


\bibliography{references}{}
\bibliographystyle{aasjournalv7}



\end{document}